\documentclass[a4paper,10pt,reqno]{amsart}
\usepackage[utf8]{inputenc}
\usepackage[%
    hyperref, backend=biber, doi=false,url=false,sorting=none,backref=true]{biblatex}
\DefineBibliographyStrings{spanish}{andothers={\textsl{et al}\adddot}}
\renewbibmacro{in:}{}    
\usepackage{csquotes}

\usepackage{amsmath}
\usepackage{amssymb}
\usepackage{amsaddr}
\usepackage{bbm}
\usepackage{color}

\usepackage{graphicx}
\usepackage{epstopdf}
\usepackage{hyperref}
\hypersetup{colorlinks=true,linktoc=page,linkcolor=blue,citecolor=red,urlcolor=cyan}

\bibliography{bibliografia}

\usepackage[english]{babel}

\newcommand{\Nf}{N_{\text{f}}}
\newcommand{\Eqref}[1]{eq.~(\ref{#1})}

\newcommand{\Tr}{\text{Tr}}
\newcommand{\xF}{x_{\text{F}}}
\newcommand{\xI}{x_{\text{I}}}
\newcommand{\yF}{y_{\text{F}}}
\newcommand{\yI}{y_{\text{I}}}
\newcommand{\nL}{n_{\text{L}}}
\newcommand{\mWS}{m_{\text{WR}}}

\newcommand{\mWSb}{m_{\bar{\text{WS}}}}

\newcommand{\ooT}{{\scriptsize \frac{1}{T}}}
\newcommand{\sqed}{S${}^2$QED}

\title{Propagator from Nonperturbative Worldline Dynamics}
\author{Sebasti\'an~Franchino-Vi\~nas}
\address{Theoretisch-Physikalisches Institut, Friedrich Schiller Universit\"at Jena, Max Wien Platz 1, 07743 Jena, Germany.}
\address{Departamento de F\'isica, Facultad de Ciencias Exactas
Universidad Nacional de La Plata, C.C.\ 67 (1900), La Plata, Argentina.}
\email{sa.franchino@uni-jena.de}
\author{Holger~Gies}
\address{Theoretisch-Physikalisches Institut, Abbe-Center of Photonics, Friedrich Schiller Universit\"at Jena, Max Wien Platz 1, 07743 Jena, Germany.}
\address{Helmholtz Institute Jena, Fröbelstieg 3, 07743 Jena, Germany.}
\email{holger.gies@uni-jena.de}

\begin{document}

\begin{abstract}

  We use the worldline representation for correlation functions
  together with numerical path integral methods to extract
  nonperturbative information about the propagator to all orders in
  the coupling in the quenched limit (small-$\Nf$ expansion). Specifically,
    we consider a  simple two-scalar field theory with cubic interaction (\sqed)  in four
  dimensions as a toy model for QED-like theories.     
  Using a worldline
    regularization technique, we are able to analyze the divergence structure
    of all-order diagrams and to perform the renormalization of the
    model nonperturbatively. Our method gives us access to a wide range of
  couplings and coordinate distances. We compute the pole mass of the
    \sqed{} electron and observe sizable nonperturbative effects in the
    strong-coupling regime arising from the full photon dressing. We also find
    indications for the existence of a critical coupling where the photon
    dressing compensates the bare mass such that the electron mass
    vanishes. The short distance behavior remains unaffected by the photon
    dressing in accordance with the power-counting structure of the model.
  
\end{abstract}

\maketitle


\section{Introduction}
In addition to the universal tool of Feynman diagram calculus for
perturbative expansions, quantum field theory has given rise to a wide
range of methods to deal with systems with many degrees of
freedom. The present work is based on the worldline method
\cite{Feynman:1950ir, Feynman:1955zz, Halpern:1977ru, Polyakov:1987ez, Bern:1987tw, Strassler:1992zr, Schmidt:1993rk, Dittrich:2000wz, Schubert:2001he}
which can be useful in both perturbative as well as nonperturbative
contexts \cite{Affleck:1981bma,Schreiber:1997jd,Brambilla:1997ky,Savkli:1999rw,SanchezGuillen:2002rz,Savkli:2004bi,BarroBergflodt:2004qa,Fosco:2004rw,Gies:2005sb,Fosco:2005bg,Dunne:2006ff, BarroBergflodt:2006hh,Bastianelli:2014bfa,Epelbaum:2015cca,Dietrich:2015oba,Ahmadiniaz:2015xoa,Gould:2017fve,Gould:2018ovk}. 

In addition to the systematic expansion in powers of a small
parameter, e.g., a coupling, the topology of diagrams is an ordering
principle of Feynman diagram calculus that is also extensively used in
modern computer-algebraic realizations, e.g. \cite{Mertig:1990an}. As a
consequence, single diagrams or single topologies may not respect the
symmetries of a theory individually, but a sum over topologies may be
needed in order to preserve a symmetry to a given order in the
parametric expansion. By contrast, the worldline formulation allows to
assess symmetry constraints already on the basis of individual
contributions. This is, because subclasses of different topologies can
be combined into a single worldline expression,
cf. \cite{Schubert:2001he,Huet:2018ksz}.

Even beyond perturbative expansion, it has already been known in the
early works on the worldline formalism \cite{Halpern:1977ru} that whole
subclasses of infinitely many Feynman diagrams can be combined into a
single closed form expression in specific theories. A prominent
example is given by scalar quantum electrodynamics (QED), where all
diagrams contributing to the one-particle irreducible (1PI) effective
action with one charged-particle loop but arbitrarily many internal
photon exchanges can be written as one worldline integral
\cite{Affleck:1981bma}. Introducing $\Nf$ flavors of charged
particles, this subclass of Feynman diagrams provides the
leading-order result of a small-$\Nf$ expansion but remains fully
nonperturbative in the gauge coupling.

Obtaining nonperturbative results from such all-order expressions is
nevertheless challenging, since their evaluation also requires a
nonperturbative way to perform the necessary renormalization, i.e.,
the fixing of physical parameters. In fact, a whole research program has been initiated to cross-check the result for the
Schwinger pair production rate evaluated from the all-order worldline
expression by semiclassical instanton methods
\cite{Affleck:1981bma} with explicit higher-loop calculations \cite{Huet:2010nt,Huet:2011kd,Huet:2017sts,Huet:2018kjj,Huet:2018ksz}, the latter requiring an explicit
treatment of the mass renormalization. If semiclassical methods turned
  out to be reliable also  at higher loop order, many studies on Schwinger
  pair production using worldline instanton methods \cite{Dunne:2006ff,Dunne:2006st,Schutzhold:2008pz,Ilderton:2014mla,Ilderton:2015lsa,Torgrimsson:2016ant,Torgrimsson:2017pzs,Akal:2018udh} could be generalized
  beyond perturbative loop expansions. Also nonperturbative variational
approximation techniques have been developed and successfully applied to
studies of bound-state properties and self-energies
\cite{Mano:1955nzi,Alexandrou:1991nc,Rosenfelder:1995bd,Rosenfelder:1995ra,Schreiber:1997jd,BarroBergflodt:2004qa,BarroBergflodt:2006hh}.
Recently, the worldline representation of field theory and its relation to
string theory has been used to propose a new way of defining UV complete field
theories \cite{Abel:2019ufz}.

In order to make progress with nonperturbative worldline techniques,
we use two ingredients as proposed in \cite{Gies:2005sb}:
first, we use a double scalar model to which 
we refer as \sqed{}. From the worldline perspective, it is structurally similar to QED, but ignores the spin of both electrons and photons.
It is su\-per-re\-normalizable in $D=4$ spacetime dimensions, such that one-loop
renormalization suffices to fix the physical parameters.
 Second, we
use numerical Monte Carlo worldline techniques \cite{Gies:2001zp,Gies:2001tj,Gies:2003cv, Gies:2006bt, Gies:2006cq,Aehlig:2011xg,Mazur:2014gta}
to nonperturbatively evaluate the worldline path integral. Whereas
previous work in this direction has concentrated on quantum effective actions or energies
\cite{Affleck:1981bma,Gies:2005sb,Gies:2006bt,Dunne:2006st,} or bound-state
amplitudes \cite{BarroBergflodt:2004qa,Savkli:2004bi, BarroBergflodt:2006hh}, we investigate the propagator of the
``charged'' scalar nonperturbatively in this work. The worldline
expression for this propagator includes all Feynman diagrams of
arbitrary (scalar) photonic self-energy corrections and thus allows to
extract information about the decay of nonperturbative correlations
with distance and the dependence on the coupling strength in this model.

While the divergence structure of the scalar model is considerably
simpler as in renormalizable models, the practical problem of
isolating and subtracting the divergent subdiagrams still persists
also in this super-renormalizable model. We show that this can be
performed with the aid of determining probability distribution
functions (PDF) for suitable building blocks of worldline
observables. For this, we generalize a technique introduced in
\cite{Gies:2005sb} to the computation of the propagator. Similar
methods have also been applied to numerical worldline computations of
Schwinger pair production \cite{Gies:2005bz} and recently to
high-accuracy results of quantum mechanical potential problems
\cite{Edwards:2017won,Edwards:2019fjh}. In the present case, the use
of a suitable analytic fit function for the PDF also allows to
extrapolate the nonperturbative results to parameter regimes where a
direct simulation is computationally expensive.

Specifically, we obtain a semi-analytical expression for the electron propagator, i.e. 
an analytical expression in which all the numerical information gathered from 
the worldline numerics is implemented through a set of parameters. 
This expression greatly simplifies the subsequent analysis, mainly focusing on the 
physical (pole) mass. We observe that the physical  mass 
gets dressed in a way that suggests the existence of a critical coupling
for which the pole mass vanishes. Moreover, there is a clear enhancement of this behaviour
arising from the all-order photon corrections in our nonperturbative expression 
compared with the one-loop result.

In the small distance regime, our nonperturbative results for the correlation
function are compatible with a vanishing anomalous dimension. This indicates
that the superrenormalizable structure of the model suggested by
power-counting also persists beyond perturbation theory.

The paper is organized as follows: in Sect. \ref{sec:worldline} we begin introducing 
 \sqed{} and deriving the analytical expression used througout
the article for the propagator of the ``charged'' scalar. The propertime discretization 
of this propagator is presented in Sect. \ref{sec:weak-coupling} as a method to regularize
the expressions and a closed formula for the one-loop contribution to the 
propagator is obtained. In possession of this analytic background, 
we show in Sect. \ref{sec:montecarlo} our numerical results. Firstly, a comparison with 
a one-loop expansion is done in Sect. \ref{sec:one-loopcomp}. Secondly, we analyse the
probability distribution function for the potential involved in our model in Sect.
\ref{sec:distribution_interaction}. Thirdly, in Sect. \ref{sec:fermion_propagator} we
study and discuss the Worldline Montecarlo results obtained for the propagator of the charged scalar.  
We state our conclusions in Sect. \ref{sec:conclusions}, while we leave an extensive analysis of the 
new $v$ lines algorithm to App. \ref{sec:vlines}. Finally, the remaining 
appendices deal in detail with the one-loop expressions (App. \ref{sec:Vcomp}), the asymptotics 
of the discretized potential (App. \ref{sec:largeN_asymptotic}), the self-energy 
in different regularizations (App. \ref{sec:additional-1-loop}) and 
the large-distance asymptotics of the one-loop propagator (App. \ref{app:largedistance}).

\section{Worldline formalism for \sqed{}}\label{sec:worldline}

Following \cite{Gies:2005sb}, we consider \sqed{}, a quantum field
theory with two interacting real scalar fields in $D$-dimensional
Euclidean spacetime with cubic interaction, as described by the Lagrangian
\begin{equation}
\mathcal{L}= \frac{1}{2} (\partial_\mu \phi)^2 + \frac{1}{2} m^2 \phi^2 + \frac{1}{2} (\partial_\mu A)^2 -\frac{i}{2} h A\phi^2.
\label{eq:Lagrangian}
\end{equation}
The model is designed to resemble QED with $A$ corresponding to the
massless photon, and $\phi$ representing the charged particle
(electron).
Apart from a global $\mathbbm{Z}_2$ symmetry for the
$\phi$ field, there is actually no local symmetry that would play a
similar role as in QED; hence the model if taken literally can be
expected to behave rather differently compared with QED. Nevertheless,
the important point for the present purpose is that the model gives
rise to a Feynman diagrammar with the same topological features as QED
-- and also has a worldline representation very similar to that of
QED.
This model is used in different versions, mostly with a real coupling, for many purposes, e.g., 
lately also for studying the decoupling in curved spaces \cite{Ribeiro:2019xgu} 
or unitarity and ghosts after the inclusion 
of higher-derivative terms \cite{Donoghue:2019fcb}.

In the present work, we are interested in the propagator, i.e. the
two-point correlator, of the $\phi$ field which can be derived 
from first principles from the generating functional
\begin{eqnarray}
Z[\eta]&=&e^{W[\eta]} = \int \mathcal{D} A \mathcal{D}\phi e^{-\int \mathcal{L} + \int \eta \phi}\nonumber\\
&=& \int \mathcal{D} A \, \det{}^{-1/2}\big(K[A]\big)\, e^{-\frac{1}{2} \int (\partial_\mu A)^2 + \frac{1}{2} \int \eta K^{-1}[A] \eta},
\label{eq:genfunc}
\end{eqnarray}
where $\eta(x)$ is an auxiliary source for the $\phi$ field, and
$K[A]=-\partial^2 +m^2 -i hA$ denotes the Klein-Gordon operator in the
background of an $A$ field. In the second line of \Eqref{eq:genfunc},
we have performed the Gau\ss ian $\phi$ integration. The $\phi$
propagator, being the connected part of the two-point function, can
straightforwardly be obtained from the Schwinger functional $W[\eta]$,
\begin{eqnarray}
  G(\xF,\xI)&=& \frac{\delta^2 W[\eta]}{\delta \eta(\xF) \delta\eta(\xI)}\bigg|_{\eta=0}\nonumber\\
  &=& \frac{1}{Z[0]} \int \mathcal{D}A\, \det{}^{-1/2} \big(K[A]\big)\, e^{-\frac{1}{2} \int (\partial_\mu A)^2}\, K^{-1}[A](\xF,\xI).
\label{eq:prop1}
\end{eqnarray}
So far, our derivation has been exact. From now on, we confine
ourselves to the leading order in a small-$\Nf$ expansion. Formally
introducing $\Nf$ flavors of the $\phi$ field, the scalar
determinant is of order $\det{}^{-1/2} \big(K[A]\big) = e^{-\frac{1}{2}
  \ln \det \big(K[A]\big)}= e^{-\frac{1}{2} \Tr \ln \big(K[A]\big)}
\sim e^{-\mathcal{O}(\Nf)}$. Thus, the determinant can be dropped to leading order
which is reminiscent to a quenched approximation. 

The worldline representation can now be introduced by rewriting the
inverse Klein-Gordon operator as a propertime integral, and
successively interpreting this operator as the Hamiltonian of a
quantum-mechanical propertime evolution. The latter is then written in
terms of a Feynman path integral,
\begin{eqnarray}
K^{-1}[A](\xF,\xI)&=&\int_0^\infty dT\, \langle \xF| e^{-K[A]T} | \xI\rangle,\nonumber\\
&=&\frac{1}{(4\pi)^{D/2}} \int_0^\infty \frac{dT}{T^{D/2}}\, e^{-m^2T} e^{- \frac{(\xF-\xI)^2}{4T}} \Big\langle e^{ih\int_0^T d\tau A(x(\tau))} \Big\rangle_{\xI}^{\xF}\!,
\label{eq:kernel1}
\end{eqnarray}
where we have introduced the worldline expectation value with respect
to the free path integral from $\xI$ to $\xF$ in propertime $T$ for an
observable $O[x]$,
\begin{equation}
\Big\langle O[x] \Big\rangle_{\xI}^{\xF} := \frac{
\int_{x(0)=\xI}^{x(T)=\xF} \mathcal{D}x\, O[x]\, e^{- \frac{1}{4} \int_0^T d\tau \dot{x}^2} 
}
{
\int_{x(0)=\xI}^{x(T)=\xF} \mathcal{D}x\,  e^{- \frac{1}{4} \int_0^T d\tau \dot{x}^2} 
}.
\label{eq:expvalue}
\end{equation}
Inserting \Eqref{eq:kernel1} into \Eqref{eq:prop1}, we can interchange
the worldline expectation value with the functional integral over $A$
fields. Introducing the current of a $\phi$ particle on the worldline,
\begin{equation}
j(z)=i h \int_0^T d \tau\, \delta^{(D)}(z-x(\tau))\quad \Rightarrow\quad e^{ih\int_0^T d\tau A(x(\tau))}= e^{\int j A},
\label{eq:phicurrent}
\end{equation}
the functional integral over the $A$-field configurations is Gau\ss{}ian, resulting in
\begin{equation}
\frac{1}{Z[0]}\int \mathcal{D}A\, e^{- \frac{1}{2} \int (\partial_\mu A)^2 + \int jA} = e^{\frac{1}{2} \int j \Delta j}.
\label{eq:Aint}
\end{equation}
Note that also the normalization $Z[0]$ has to be computed within the
small-$\Nf$ approximation. In \Eqref{eq:Aint},
$\Delta=(-\partial^2)^{-1}$ denotes the propagator of the photonic $A$
field which  in coordinate space reads
\begin{equation}
\Delta(x_1,x_2)= \frac{\Gamma\left( \frac{D-2}{2} \right)}{4 \pi^{D/2}} \frac{1}{|x_1-x_2|^{D-2}}.
\label{eq:Delta}
\end{equation}
We observe that the photon fluctuations can be summarized in a
current-current interaction of the $\phi$ field being represented by a
worldline trajectory with itself. Using
the explicit representation of the current \eqref{eq:phicurrent}, we
get
\begin{eqnarray}
\frac{1}{2}\int j\Delta j &=& - \frac{h^2}{8\pi^{D/2}} \Gamma\left({\scriptstyle \frac{D-2}{2} }\right) \int_0^T d\tau_1 d\tau_2 \frac{1}{|x_1-x_2|^{D-2}} \nonumber\\
&=:& - g V[x], \quad g:= \frac{h^2}{8 \pi^{D/2}} \Gamma\left({\scriptstyle \frac{D-2}{2} }\right).
\label{eq:V1}
\end{eqnarray}
Here and in the following, we use the convention $x_i=x(\tau_i)$. The
coupling $g$ plays the role of a fine-structure constant in this
model, and $V[x]$ can be viewed as a potential for the interactions of
a worldline with itself. Inserting these findings into
\Eqref{eq:prop1}, we end up with the worldline representation of the
(unrenormalized) field propagator to leading order in the small-$\Nf$
limit,
\begin{equation}
  G(\xF,\xI)=\frac{1}{(4\pi)^{D/2}} \int_0^\infty \frac{dT}{T^{D/2}}\, e^{-m^2T} \, e^{-\frac{(\xF-\xI)^2}{4T}} \Big\langle e^{-gV[x]}\Big\rangle_{\xI}^{\xF}, 
\label{eq:prop2}
\end{equation}
in agreement with the formula given in \cite{Gies:2005sb}. It
is obvious that this representation is nonperturbative in the coupling
$g$ as it contains powers of $g$ to all orders. In a Feynman-diagram
language, \Eqref{eq:prop2} summarizes all possible diagrams with one
$\phi$-particle line and an arbitrary number of photonic $A$-field
radiative corrections in one single expression. There is no
restriction on the diagram topology (e.g. 1PI or ``rainbow''
diagrams): the expression also includes one-particle reducible as well
as crossing-rainbow diagrams. The  challenge pursued in the
following sections is to evaluate the worldline path integral.

\section{Weak-coupling expansion and renormalization}\label{sec:weak-coupling}

Let us start with the noninteracting limit $g\to 0$. In that case,
there is no photonic contribution, and the propagator of
\Eqref{eq:prop2} reduces to the free Green's function of the massive
Klein-Gordon operator. Upon performing the $T$ integral, we arrive at
an expression in terms of a Macdonald function,
\begin{equation}
 G_0(\Delta x)=\frac{1}{(2\pi)^{D/2}} \left(\frac{m^2}{\Delta x^2} \right)^{(D-2)/4} K_{(D-2)/2}(m\Delta x), \quad \Delta x = |\xF-\xI|.
\label{eq:propfree}
\end{equation}
Because of translational invariance, the propagator depends only on
the distance of the endpoints also in the presence of photonic
interactions. 

At weak-coupling, \Eqref{eq:prop2} suggests a perturbative expansion
in powers of the coupling, resulting in higher-order worldline
correlators of the interaction potential, $ (-g)^n \big\langle
V^n[x]\big\rangle_{\xI}^{\xF}$. Because of the superrenormalizable
structure of the theory, we expect the divergence relevant for mass
renormalization to appear only to leading order in the coupling
$g$. Once this one-loop order is renormalized, all remaining terms
should be finite. Thus, a careful analysis of the leading order
expectation value $\big\langle V[x]\big\rangle_{\xI}^{\xF}$ is a
crucial building block for the nonperturbative study. This expression
can be studied straightforwardly with continuum worldline techniques
\cite{Schubert:2001he}, allowing to make contact with standard
regularization schemes such as propertime or dimensional
regularization. 

In order to make direct contact with the full numerical studies below,
we proceed differently and study this expectation value by regulating
the path integral in terms of a propertime lattice. For this, we first
perform a rescaling of the worldlines, such that their distribution
becomes independent of the propertime \cite{Gies:2001zp}
\begin{equation}
y(t):= \frac{1}{\sqrt{T}} x(Tt),\quad t\in[0,1] \quad \Rightarrow\quad
\int_0^T d\tau\, \dot{x}^2(\tau) = \int_0^1 dt\, \dot{y}^2(t).
\label{eq:unitlines}
\end{equation}
We call the worldlines parametrized by $y(t)$ {\em unit
  lines}. Accordingly, the initial and final points of the unit lines
are related to the physical initial and final point by rescaling,
$\yI=y(0)=\xI/\sqrt{T}$, $\yF=y(1)=\xF/\sqrt{T}$. Moreover, the
worldline interaction potential can be rescaled,
\begin{equation}
  V[x]=\int_0^T \frac{d\tau_1 d\tau_2}{|x_1-x_2|^{D-2}} = T^{3-D/2} \int_0^1 \frac{ dt_1 dt_2}{|y_1-y_2|^{D-2}}, \quad
  \big\langle V[x]\big\rangle_{\xI}^{\xF} \equiv \big\langle V[y]\big\rangle_{\yI}^{\yF}.
\label{eq:rescV}
\end{equation}
Subsequently, we discretize the unit lines, by slicing the rescaled
propertime $t$ into $N$ time intervals,
\begin{equation}
y(t)\quad \to \quad y(t_i)=y_i\in\mathbbm{R}^D, \quad t_i=\frac{i}{N}, \quad i,=0,1,\dots, N.
\label{eq:ydisc}
\end{equation}
Note that the spacetime remains continuous in this approach. We
discretize the worldline kinetic term in \Eqref{eq:unitlines} by a
standard nearest-neighbor difference quotient (see App. \ref{sec:vlines}).
The integrals in the worldline
interaction potential, in the discretized form, then 
correspond to Riemann sums over the $N$ propertime slices
\begin{equation}
V[y]
=\frac{2 T^{3-D/2}}{N^2} \sum_{l=0}^{N-2}\sum_{n=l+1}^{N-1} \frac{1}{|y_l-y_n|^{D-2}}.
\label{eq:Vdisc}
\end{equation}
Here, we have removed the coincident points for $l=n$, where the
self-interaction potential in the discretized version is
ill-defined. The associated short-distance singularity will
nevertheless be approached in the propertime continuum limit for
increasing $N$, as the discretized version of the probability
distribution \eqref{eq:unitlines} corresponds to a random walk for
which $|y_i-y_{i-1}|\sim 1/\sqrt{N}$. Hence we expect a divergence to
appear in the limit $N\to \infty$. We have checked that a
regularization of the coincident point limit by shifting the
denominator $|y_l-y_n|^{D-2}\to |y_l-y_n|^{D-2}+ \delta$ but including
the $n=l$ terms in the sum yields the same divergences in the limit
$\delta\to 0$.

The procedure of keeping $N$ finite hence regularizes the worldline
expressions. As shown in the following, it allows for meaningful numerical
computations of relevant quantities as well as for a controlled study of the
$N\to\infty$ limit. It is however important to note that there is a
decisive difference to a conventional momentum-space or short-distance
cutoff: as $\delta y \sim 1/\sqrt{N}$, we have $\Delta x \sim
\sqrt{T}/\sqrt{N}$. For a given fixed $N$, fluctuations associated with
different propertimes $T$ are thus regularized at different length scales. As
a consequence, our final results will differ from those obtained by a
standard regularization scheme not only by a simple scheme change, i.e.,
a shift of finite constants. Instead, our \textit{worldline regularization}
will be linked to a standard regularization by a finite spacetime- or 
momentum-dependent transformation. The connection is worked out in
App.~\ref{sec:additional-1-loop} on the one-loop level. For reasons of
clarity, we use the worldline regularization in the main text for our
nonperturbative analysis.

Coming back to the expectation value of the worldline interaction potential, it has to be
computed for an ensemble of open worldlines interconnecting $\yI$ and
$\yF$ with a Gau\ss{}ian velocity distribution. For our numerical
studies below, we use a new algorithm (which we call $v$ lines) to generate such discretized
random paths. This construction is detailed in
App. \ref{sec:vlines} and is tested considering a simple model in App. \ref{sec:test-vlines}.

For the one-loop contribution to the propagator in the discretized
formulation, we now need to compute
\begin{equation} 
 \left\langle V[y] \right\rangle^{\yF}_{\yI}=\mathcal{N} \int_{\yI}^{\yF}\mathcal{D}y\, e^{-\frac{N}{4} \sum_{i=1}^{i=N} (y_i-y_{i-1})^2} V[y],
\end{equation}
with the discretized representation \eqref{eq:Vdisc} of the
interaction potential, and the abbreviations,
\begin{equation}
\mathcal{N}^{-1}:={\int^{\yF}_{\yI}\mathcal{D}y\, e^{-\frac{N}{4} \sum_{i=1}^{i=N} (y_i-y_{i-1})^2}}, 
\quad \mathcal{D}y := \prod_{j=1}^{N-1} d^Dy_j.
\end{equation}
The worldline integrations can be performed analytically by using the
Fourier representation of the interaction kernel, cf. \Eqref{eq:Delta},
\begin{equation}\label{eq:<V>}
\left\langle V[y] \right\rangle^{\yF}_{\yI}=\frac{8 T^{3-D/2}}{N^2 \Gamma \left( \frac{D-2}{2}\right)} \sum_{0=l<n}^{N-1} \mathcal{N}\int_{\yI}^{\yF}\mathcal{D}y\, e^{-\frac{N}{4} \sum_{i} (y_i-y_{i-1})^2} \int \frac{d^Dp}{(4\pi)^{D/2}} \frac{e^{ip(y_l-y_n)}}{p^2}. 
\end{equation}
We observe that all $y_i$ integrals still remain Gau\ss{}ian and hence
can be performed by a suitable completion of the square. The
corresponding computation can be performed along the same lines as for
the construction of the $v$ lines algorithm, except for a different
completion of the squares at positions $l$ and $n$. With the details 
highlighted in App. \ref{sec:Vcomp}, we arrive at
\begin{eqnarray} 
\left\langle V[y] \right\rangle^{\yF}_{\yI}
&=&2 \frac{T^{3-D/2}}{(\Delta y)^{D-2}} \sum_{n=1}^{N-1} \frac{N^{D-4} (N-n)}{n^{D-2}} 
\left[ 1- \frac{\Gamma\left( \frac{D-2}{2}, \frac{n \Delta y^2}{4 (N-n)} \right)}
{\Gamma\left( \frac{D-2}{2} \right)} \right], \label{eq:VdiscD}
\end{eqnarray}
where $\Delta y = |\yF-\yI|$, and $\Gamma(a,z)$ denotes the incomplete gamma function. Let us from now
on specialize to the case of $4$-dimensional spacetime, $D=4$, yielding
\begin{eqnarray}
\left\langle V[y] \right\rangle^{\yF}_{\yI}&=&2 \frac{T}{\Delta y^{2}} \sum_{n=1}^{N-1} \frac{ (N-n)}{n^{2}} 
\left[ 1- e^{- \frac{n \Delta y^2}{4 (N-n)}} \right] \label{eq:VdiscD4}\\
&=& 2 \frac{T^2}{\Delta x^2} \sum_{n=1}^{N-1} \frac{ (N-n)}{n^{2}} 
\left[ 1- e^{- \frac{n \Delta x^2}{4 T(N-n)}} \right],
\end{eqnarray}
where we reinstated the dimensionful physical distance in the last line. It is instructive to study the short-distance limit,
\begin{eqnarray}
\left\langle V[y] \right\rangle_{\yI}^{\yF}&=& \frac{1}{2} T \sum_{n=1}^{N-1} \frac{1}{n} + \mathcal{O}(\Delta y^2) \nonumber\\
&=&\frac{1}{2} T \, H_{N-1} + \mathcal{O}(\Delta y^2) \label{eq:divN}\\
&=& \frac{1}{2} T\, \big( \ln N +\gamma \big) + \mathcal{O}(\Delta y^2, 1/N). \nonumber
\end{eqnarray}
We observe that the short-distance limit is essentially given by the
harmonic number $H_{N-1}$ which diverges logarithmically for large $N$
with the constant part given by the Euler-Mascheroni constant
$\gamma$. The analytical result \eqref{eq:divN} for the large-N expectation
value of the worldine interaction potential for closed worldlines,
i.e., $\Delta y=0$, is in fact in good agreement with
the numerical estimates of \cite{Gies:2005sb}\footnote{Note that there
  is a typo in Eq.~(41) of \cite{Gies:2005sb}, where the logarithm
  $\ln$ should read $\log_2$.}.

The logarithmic divergence discovered in \Eqref{eq:divN} is a UV
divergence and indicates the necessity of performing a renormalization
of physical parameters. In the present model, simple power-counting
tells us that such a divergence is expected to correspond to a
counterterm $\delta m^2$, denoting the additive renormalization of the
mass (a possible wave function renormalization $Z_\phi$ remains
finite). In order to illustrate how the divergence relates to mass
renormalization, we go back to \Eqref{eq:prop2} and concentrate on 
two factors of the propertime integrand in the short distance limit:
\begin{eqnarray}
e^{-m^2 T} \Big\langle e^{-gV[x]}\Big\rangle_{\xI}^{\xF}&=&
e^{-m^2 T} \sum_{k=0}^\infty \frac{(-g)^k}{k!} \big\langle V[x]^k\big\rangle_{\xI}^{\xF} \nonumber\\
&=&
e^{-m^2 T} \sum_{k=0}^\infty \frac{(-g)^k}{k!} \left(\big\langle V[x]\big\rangle_{\xI}^{\xF}\right)^k + \text{connected} \nonumber\\
&=&
e^{-m^2 T} e^{-g\langle V[x]\rangle_{\xI}^{\xF}} + \text{connected}. \label{eq:conn1}
\end{eqnarray}
Here ``connected'' denotes the connected parts of operator products
of the self-interaction potential; e.g., to second order in $V$, the
connected part would essentially be given by $\langle V[x]
V[x]\big\rangle_{\xI}^{\xF}-\left(\langle
  V[x]\big\rangle_{\xI}^{\xF}\right)^2$. These connected parts
contribute to 1PI diagrams with overlapping photon radiative
corrections, which are power-counting finite. Therefore, the mass
shift is fully contained in the disconnected part written explicitly
in \Eqref{eq:conn1}. This suggests to define the following finite mass parameter
\begin{equation}
\mWS^2:= m^2 -\delta m^2, \quad \delta m^2=- \frac{g}{2} \big( \ln N + \gamma \big),
\label{eq:mWS}
\end{equation}
where we understand the bare mass $m$ to be implicitly $N$ dependent,
such that $\mWS$ is kept at a finite fixed value in the limit
$N\to\infty$. In the following, $\mWS$ will serve as a fixed input
parameter representing a finite mass parameter of the theory in the
worldline regularization. Moreover, we use $\mWS$ in the remainder of this work to
set the scale for all other dimensionful quantities. 

To sum up, we conclude that
the mass-renormalized version of the propagator \eqref{eq:prop2} in
worldline representation and for $D=4$ reads,
\begin{equation}
  G(\xF,\xI)=\frac{1}{(4\pi)^{2}} \int_0^\infty \frac{dT}{T^{2}}\, e^{-\mWS^2T} \, e^{-\frac{(\xF-\xI)^2}{4T}} 
\Big\langle e^{-\delta m^2 T-gV[x]}\Big\rangle_{\xI}^{\xF}, 
\label{eq:prop3}
\end{equation}
yielding a finite result also in the limit $N\to\infty$, because of
\Eqref{eq:divN}.

\section{Worldline Monte Carlo for the propagator}\label{sec:montecarlo}

Let us now turn to an evaluation of the worldline path integrals 
by a Monte Carlo procedure. For this, we generate an ensemble of 
open worldlines extending over the dimensionless distance 
$\Delta y=\yF-\yI$ using the $v$ lines algorithm. 
This ensemble is characterized by the total number $\nL$ of 
lines and the number $N$ of discretization points per line. 
The worldline expectation value of an operator $\mathcal{O}[y]$ is then estimated 
as\footnote{Here and in the following, we use the convention that the symbol $\simeq$ relates 
the quantities computed analytically to the corresponding ones obtained using WMC.}
\begin{equation}
\left\langle \mathcal{O}[y] \right\rangle^{\yF}_{\yI}\simeq \frac{1}{\nL} \sum_{\ell=1}^{\nL} \mathcal{O}[y_\ell].
\label{eq:MC}
\end{equation}
The full continuum result would be obtained in the limit $\nL\to \infty$ and $N\to\infty$.

\subsection{One-loop comparison}
\label{sec:one-loopcomp}

As a benchmark test, we compare the results of a Monte Carlo simulation for
the expectation value of the interaction potential with the analytical results, 
cf. \Eqref{eq:VdiscD4}; this corresponds to a check at the one-loop level. 
For this purpose, we generate ensembles of $\nL=100000$ lines for an increasing 
number $N$ of points per line extending over a distance 
$\Delta y=\left\vert y_F-y_I\right\vert$. 
In the following, we set the propertime $T=1$ without loss of generality; 
for dimensional reasons, it is clear that $V[y]$ scales linearly with $T$, cf. \Eqref{eq:rescV}.

\begin{figure}[h]
\begin{minipage}{.48\textwidth}
\includegraphics[width=1.05\textwidth]{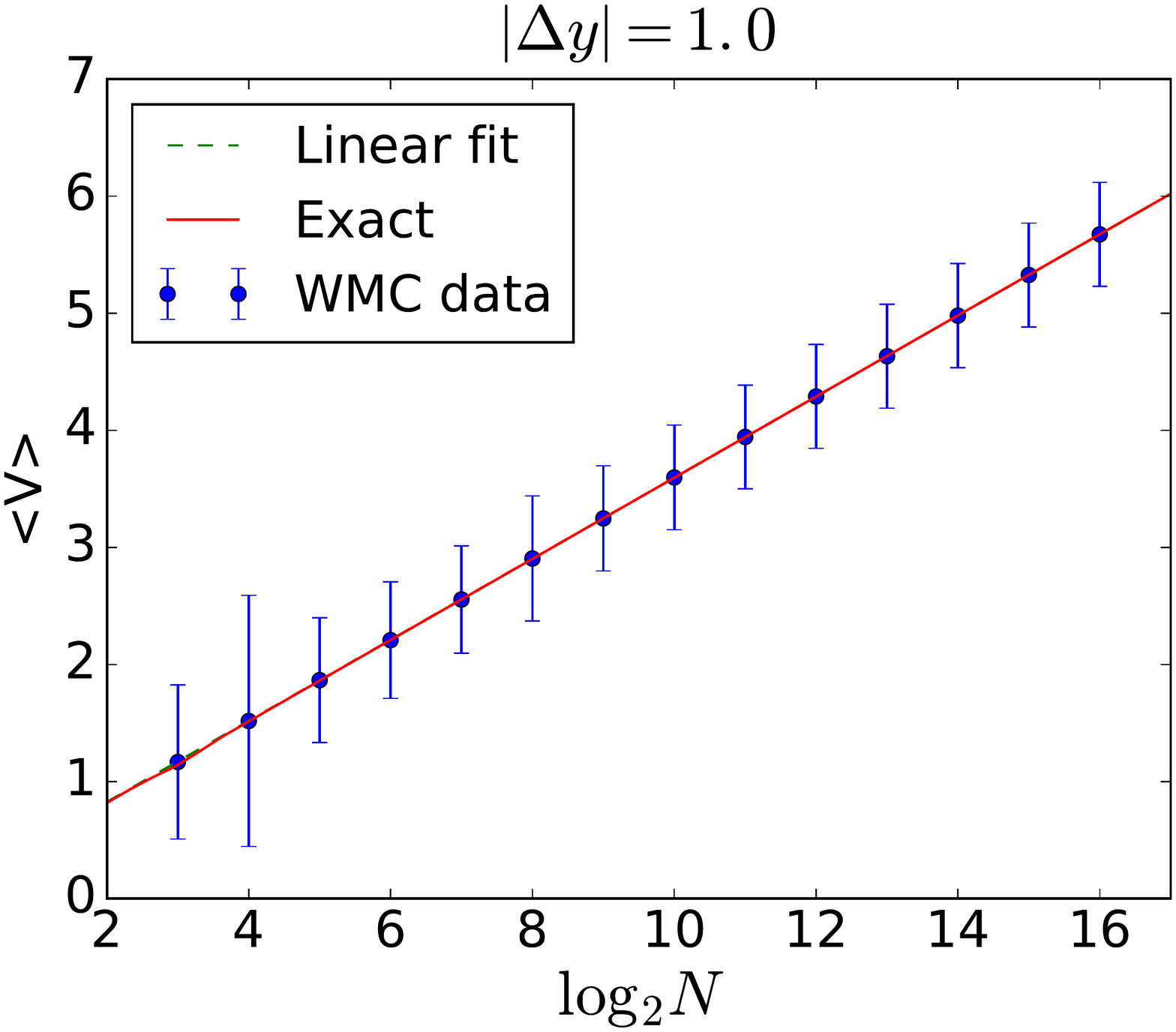}
\end{minipage}
\begin{minipage}{.48\textwidth}
\includegraphics[width=1.05\textwidth]{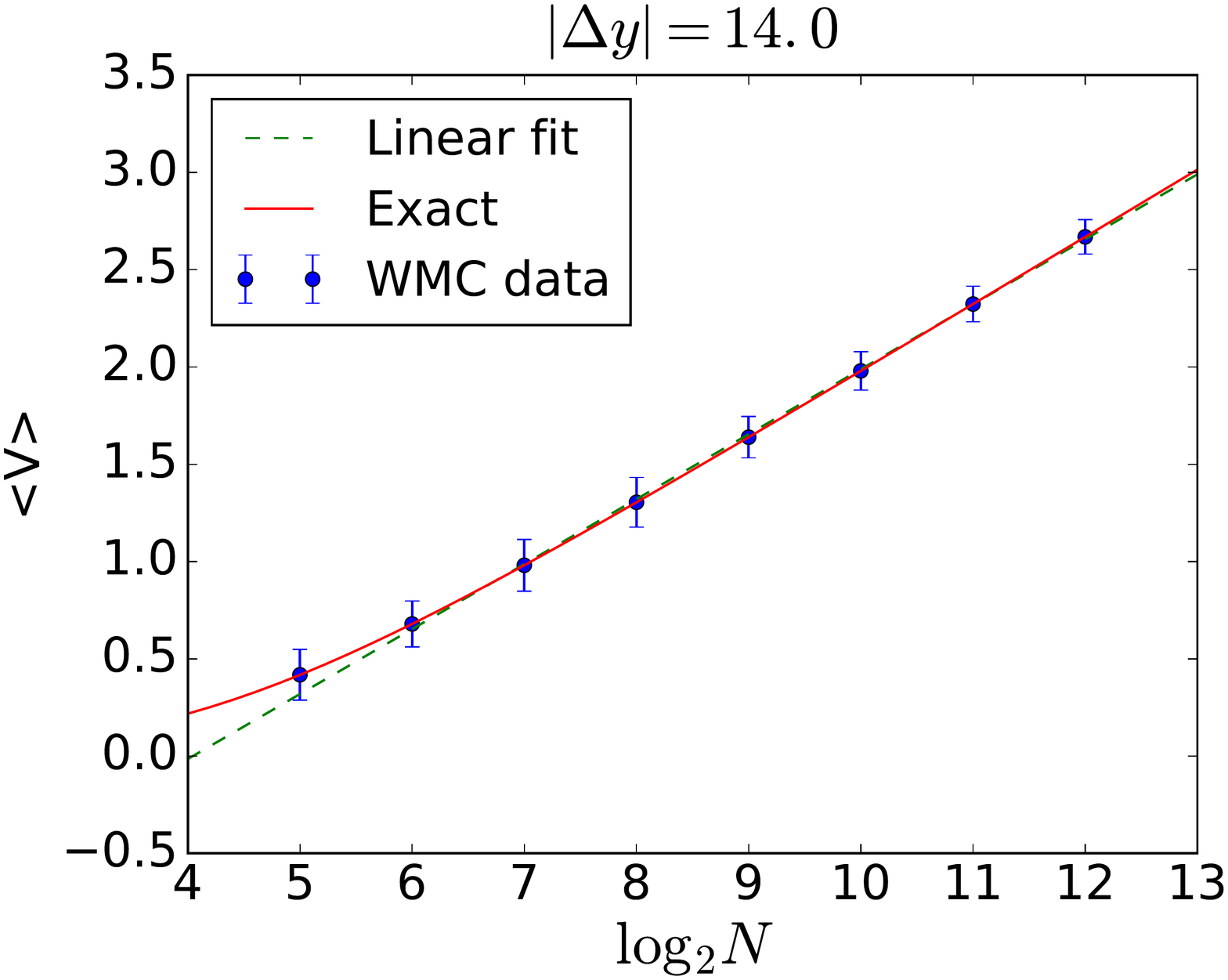}
\end{minipage}
\caption{\footnotesize Mean value $\langle V\rangle$ of the potential as a function of $\log_2 N$, the base two logarithm of the number $N$ of point per loops, for lines whose endpoints distance are $\Delta y= 1$ (left) and $\Delta y=14$ (right).  The exact analytic expression (solid red line) and the fitting with a straight line (dashed green line) are also shown.}
\label{fig.one_loop_x1}
\end{figure}

In Fig.~\ref{fig.one_loop_x1}, we show the behavior of the expectation value of the potential $\langle V \rangle_{\yI}^{\yF}$ as a function of the number of points per line $N$ (in a $\log_2$ scale); the left (right) panel corresponds to $\Delta y=1$ ($\Delta y=14$). The error bars for the numerical data correspond to the root-mean-square (RMS) of the expectation value in the given ensemble. The worldline Monte Carlo (WMC) data is compared to the analytical expression of \Eqref{eq:VdiscD4} (red solid line).
The agreement between the numerical data and the analytical result is
very satisfactory for all $N$ and $\Delta y$. The RMS error appears to
overestimate the true error. For larger $\Delta y$, the error becomes
slightly smaller, as the large distance between the end points forces
the lines to be closer to the classical paths with less fluctuations.

We also show a least squares fit to the data 
(green dashed line) using a fit function linear in $\log_2 N$,
\begin{equation}
f_{\text{fit}}= a_V \log_2 N + b_V,
\label{eq:largeNfit}
\end{equation}
confirming the presence of the logarithmic divergence with the
regularization parameter, also found analytically in
\Eqref{eq:divN}. For larger $\Delta y$ (right panel of
Fig.~\ref{fig.one_loop_x1}), we observe slight differences between the
data/exact results and the linear fit in $\log_2 N$, indicating the
onset of the higher order terms in \Eqref{eq:divN}.

\begin{figure}[h]
\begin{minipage}{.48\textwidth}
\includegraphics[width=1.05\textwidth]{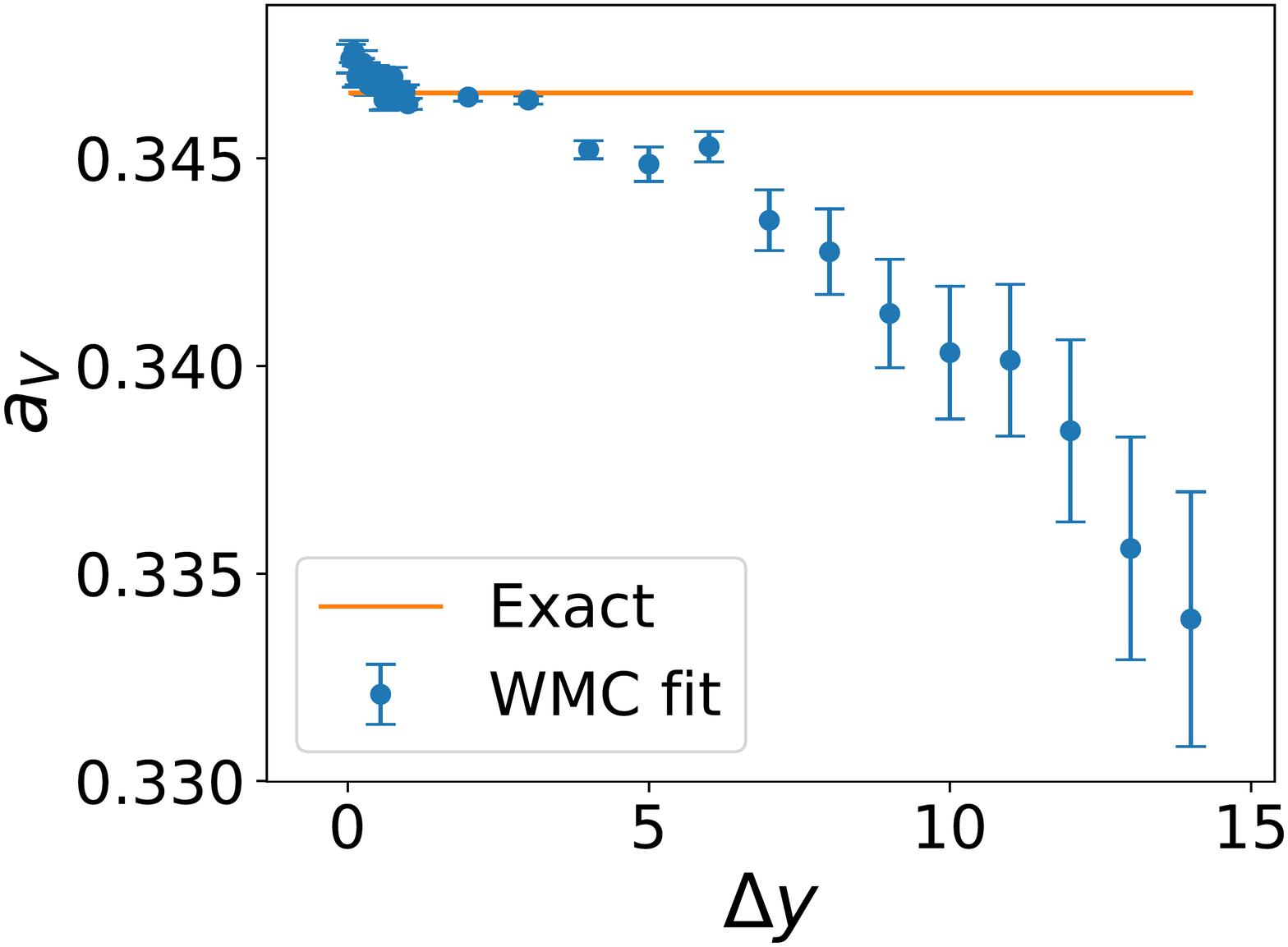}
\end{minipage}
\begin{minipage}{.48\textwidth}
\includegraphics[width=1.05\textwidth]{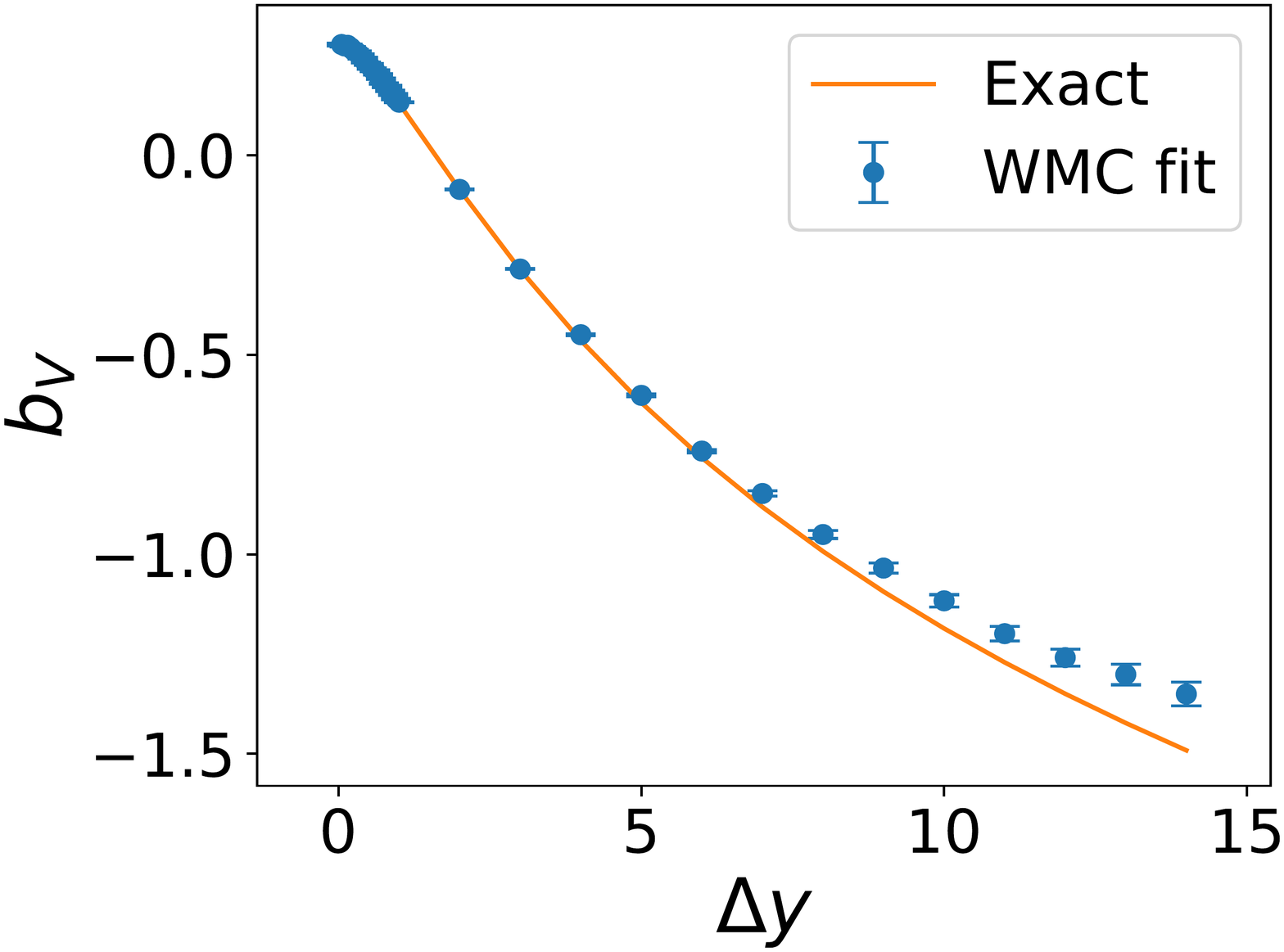}
\end{minipage}
\caption{\footnotesize Fit parameters $a_V$ (left) and $b_V$ (right) of the
  fit \Eqref{eq:largeNfit} to the expectation value of the interaction
  potential (light blue points), as functions of the distance $\Delta y$. The orange solid
  lines show the exact analytical prefactor $(\ln 2)/2$ of the log divergence 
  (left) and the remainder function of \Eqref{eq:VdiscD4} after
  subtraction of the log-divergence (right).} 
\label{fig.one_loop_linear_fit}
\end{figure}

To illustrate this point more quantitatively, we show the results for
the fit parameters $a_V$ and $b_V$ as a function of increasing
$\Delta y$ in Fig.~\ref{fig.one_loop_linear_fit}. Whereas for small
values of $\Delta y$ the fit parameter for $a_V$ settles near the
exact value $\frac{\ln 2}{2} \simeq 0.3466$ for the large-$N$ limit 
(left panel), we observe deviations on the few-percent level kicking
in for $\Delta y \gtrsim 5$. This implies that a larger range of $N$
points per loops is required to isolate the $\ln N$ divergence on the,
say, 1\% level for larger distances $\Delta y$. 

This is also visible for the terms of order $N^0$: the right panel in
Fig.~\ref{fig.one_loop_linear_fit} shows the result for the parameter
$b_V$ starting off near the analytical result $\gamma/2\simeq 0.2886$
for small $\Delta y$ and becoming negative for larger $\Delta y$. Our
simulation data is satifactorily close to the exact $\Delta
y$-dependence as predicted by \Eqref{eq:VdiscD4}. Again, the
deviations kicking in for larger $\Delta y \gtrsim 6$ are indicative
for the fact that $N$ has not been chosen sufficiently big in the
simulations in order to isolate the logarithmic divergence in $N$ from
the $\Delta y$-dependent finite terms.

It is useful to turn this observation around: for a given finite
$\Delta y$, we may ask how many points per line $N$ are needed to
reliably determine the logarithmic divergence as is required for a
proper renormalization. Demanding for a certain precision for this
procedure, we can obtain an estimate for a minimal number of $N$ to
have simulational access to the one-loop structure of the
system. Below, we call this procedure the ``one-loop test''.

\subsection{Probability distribution for the interaction potential}\label{sec:distribution_interaction}

With this validation of our numerical methods, also showing the
necessity to go to large $N$ for some quantities, we now need a method to deal with expectation values of functions of the
interaction potential
%
$
\left\langle f(V[x])\right\rangle_{\xI}^{\xF},
$
%
 cf. \Eqref{eq:prop3},
An obvious difficulty is the isolation of the logarithmic divergencies for increasing $N$. 
A naive subtraction of the analytically known divergence in \Eqref{eq:prop3} would be problematic, 
as any numerical error in determining this divergence 
gets amplified with $\ln N$ after analytical subtraction. 

In the following, we use the method of numerically determining the
probability distribution of the relevant observable as introduced in
\cite{Gies:2005sb} for the coincidence limit and generalize it to
finite distances $\Delta x$. We define the distance-dependent
probability distribution function (PDF) $\mathcal{P}(v,\Delta y)$ for
the potential 
\begin{align}
\begin{split}
 \mathcal{P}(v,\Delta y)=\frac{\int_{y(0)=\yI}^{y(1)=\yF}\mathcal{D}y\, e^{- \int_0^1 \frac{\dot{y}^2}{4}} \delta\left(\ooT V[y]-v\right) }{\int_{y(0)=\yI}^{y(1)=\yF}\mathcal{D}y\, e^{- \int_0^1 \frac{\dot{y}^2}{4}} },
\end{split}
\end{align}
where we have scaled out the trivial linear propertime dependence of
$V[y]$, cf. \Eqref{eq:rescV}, such that $ \mathcal{P}(v,\Delta y)$ is
not explicitly $T$-dependent.  In addition to $\Delta y$, the PDF using a discretized definition also depends on
$N$. Once the PDF is known, the desired expectation value can be
computed from
\begin{align}
\left\langle f\left( V[y]\right)\right\rangle_{y_I}^{y_F}=\int_0^{\infty} dv\, \mathcal{P}(v, \Delta y)\, f( T v).
\end{align}
In the following, we determine the PDF from numerical data, 
i.e., from binned histograms of the observable $V[y]$, and analyze it 
with a suitable fit function. For the latter, we choose
\begin{align}\label{pdf}
 P(v,\Delta y):=\frac{\beta^{1+\alpha}}{\Gamma(\alpha+1)} (v-v_0)^{\alpha} e^{-\beta(v-v_0)}\theta(v-v_0),
\end{align}
where the fit parameters $\alpha$, $\beta$ and $v_0$ are all $\Delta
y$ dependent\footnote{
A more elaborate choice for the coincidence limit has been made in 
\cite{Gies:2005sb}; beyond the coincidence limit, we find that the 
present simpler choice appears more suitable.}.
The fit function is already normalized, $\int_0^{\infty} dv\,
P(v,\Delta y)=1$. Also, it has been designed in such a way that the
$N$-dependent log divergence can be carried by the parameter $v_0$,
see below.

In Fig.~\ref{fig.distribution} (left panel), we depict the numerically
obtained PDF $\mathcal{P}(v,\Delta y)$ using 100 bins for the case of
a closed loop (coincidence limit with $\Delta y=0$) and $N=2^{5}$
points per loop. 
\begin{figure}[t] 
\begin{minipage}{.48\textwidth}
\includegraphics[width=1.05\textwidth]{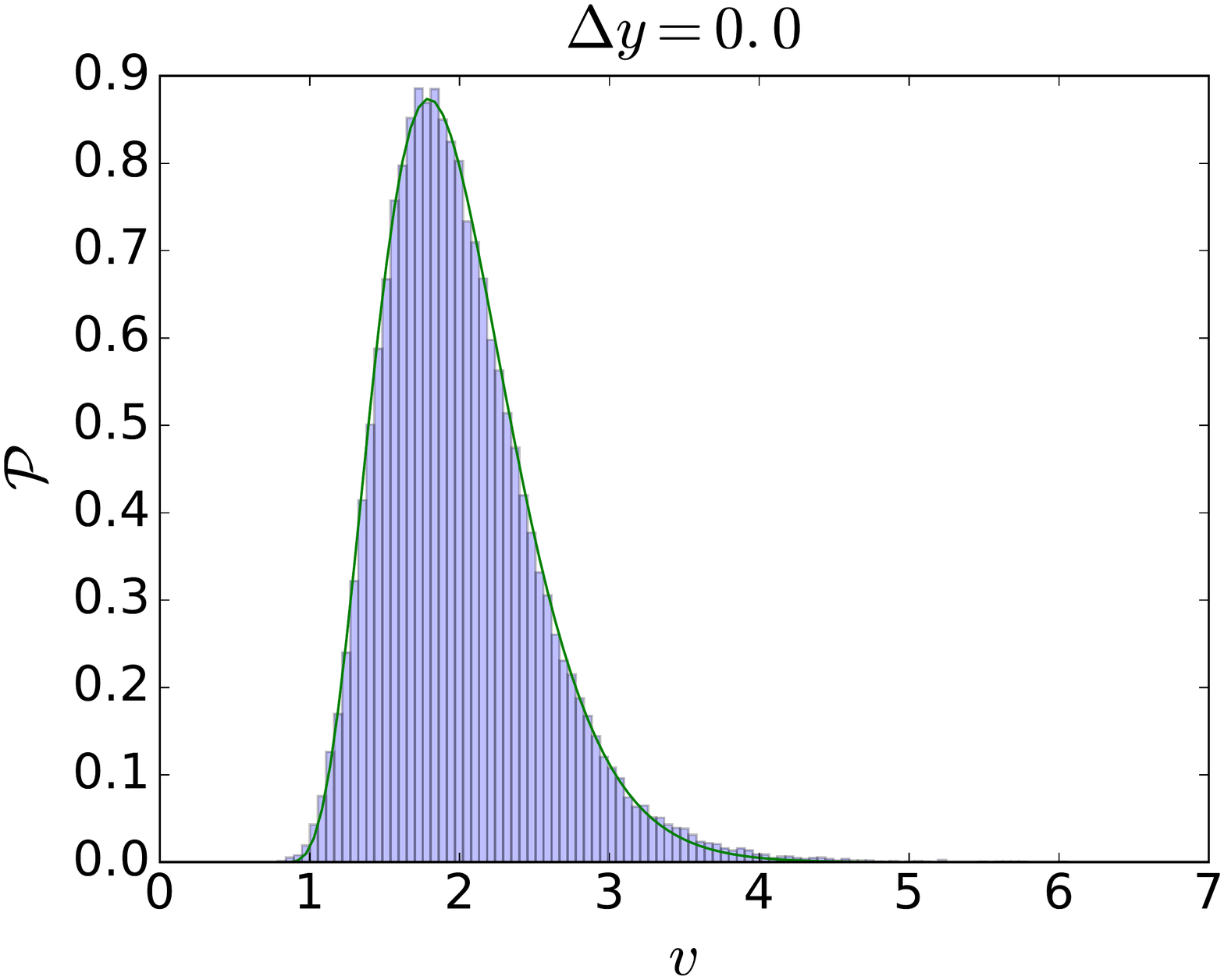}
\end{minipage}
\begin{minipage}{.48\textwidth}
\includegraphics[width=1.05\textwidth]{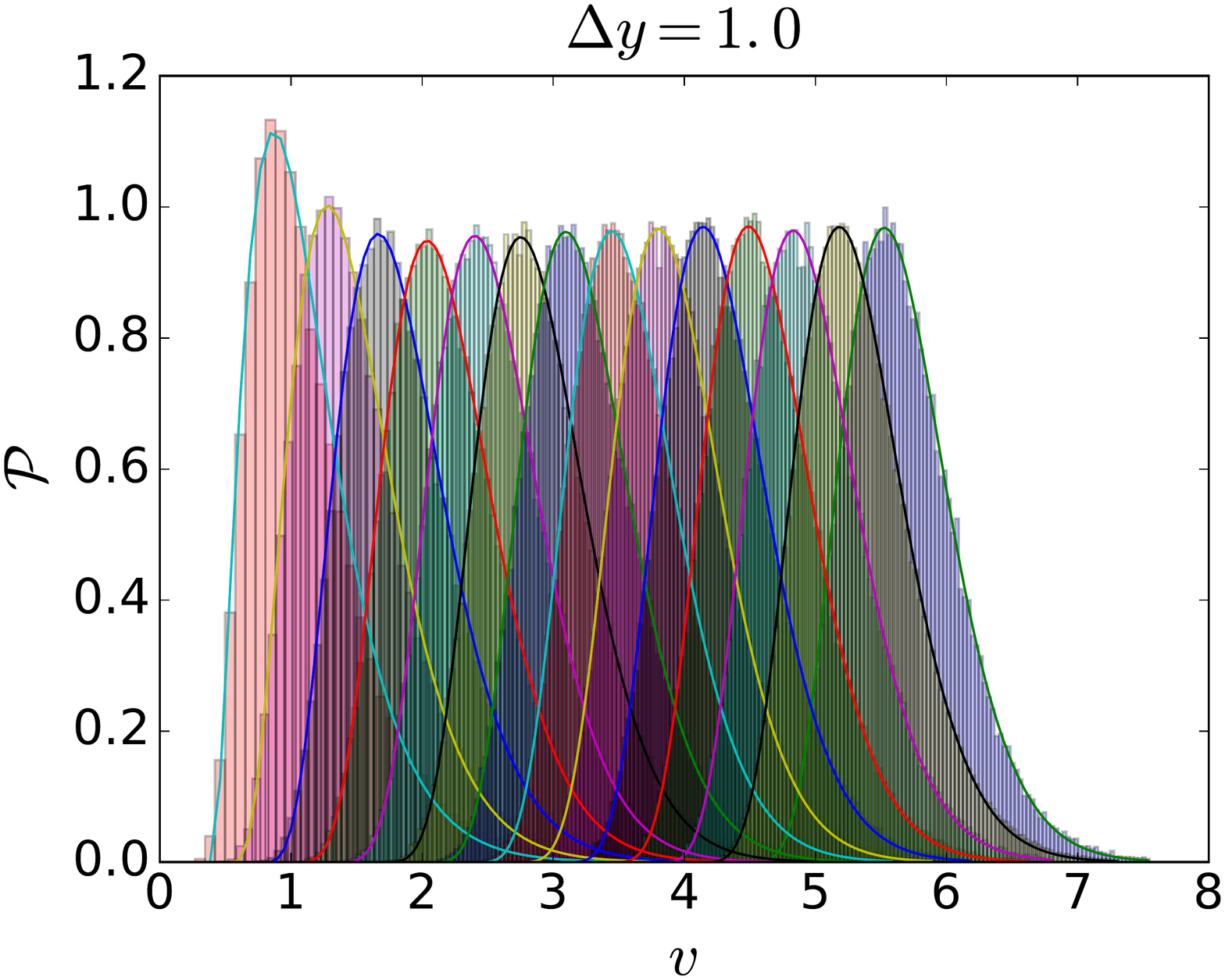}
\end{minipage}
\caption{\footnotesize Probability distributions $\mathcal{P}$ and the
  corresponding fits $P$ for $\Delta y=0$, $N=5$ (left) and $\Delta y
  =1$, $N=3,4,\cdots, 16$ (right), with $N$ increasing from left to right in the right plot.}
\label{fig.distribution}
\end{figure}
The fit $P(v,0)$ according to the ansatz \eqref{pdf}
is also shown. As is visible from the plot, the proposed fit function
$P$ is compatible with the main features of the numerical data
$\mathcal{P}$: its decay for large and small potential, the
existence of a maximum, and the position of the latter\footnote{At close inspection,
  the numerical data is slightly larger than the fit function for
  large values of $v$. The alternative fit function given in
  \cite{Gies:2005sb} would cure this feature. However, we observe in the following
  that this issue is less relevant for finite $\Delta y$.}.

\begin{figure}[h]
\begin{minipage}{.48\textwidth}
\includegraphics[width=1.05\textwidth]{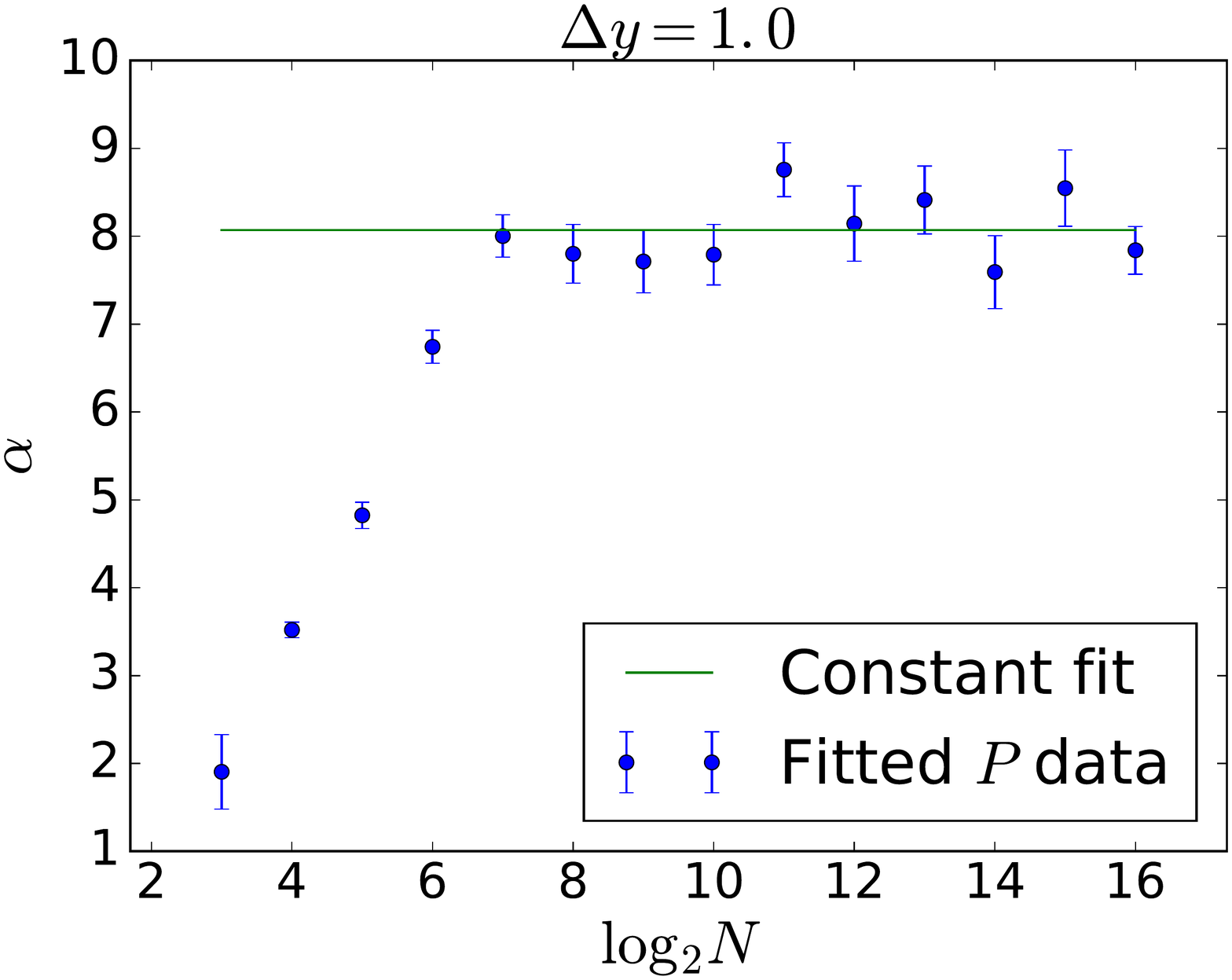}
\end{minipage}
\begin{minipage}{.48\textwidth}
\includegraphics[width=1.05\textwidth]{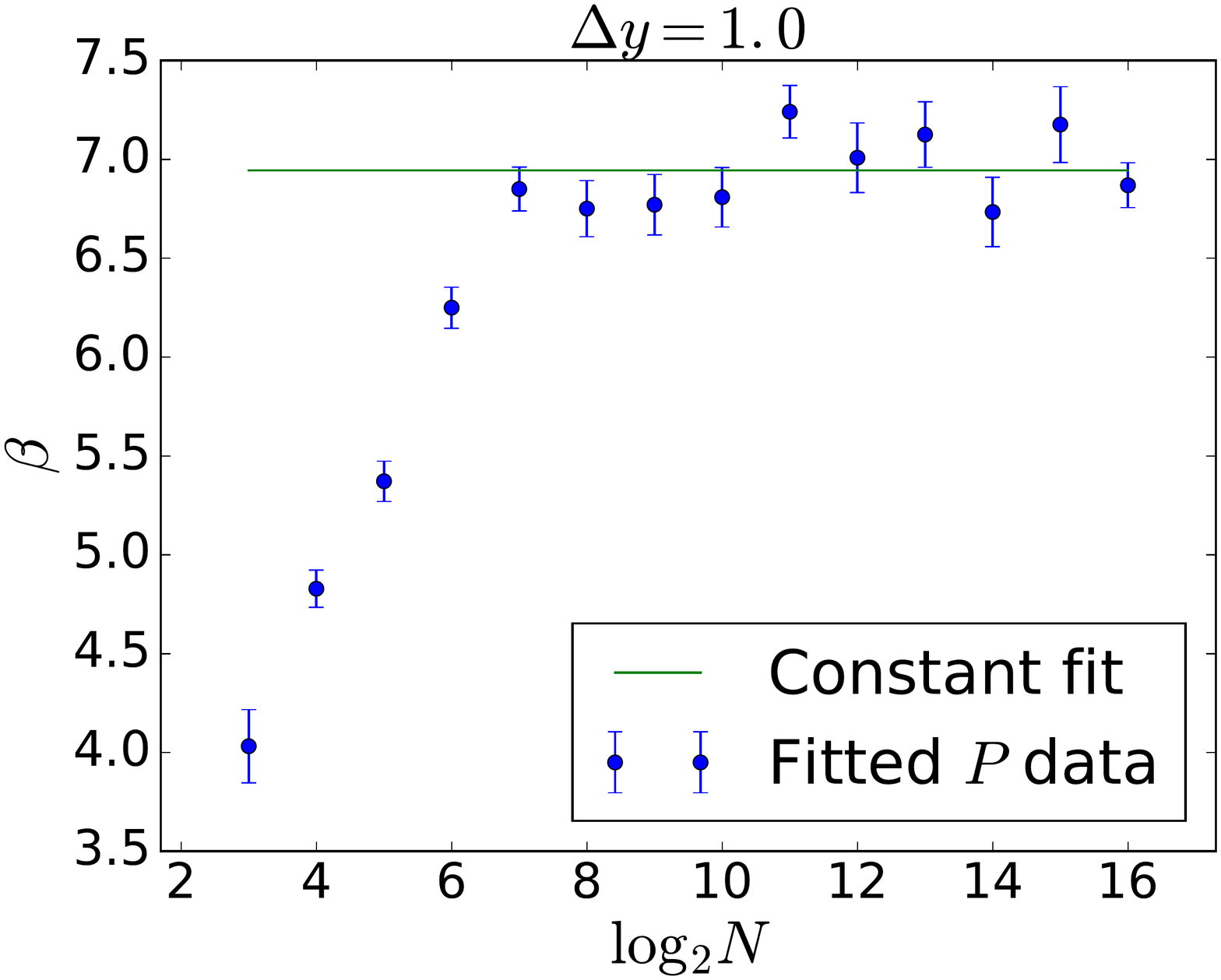}
\end{minipage}
\caption{\footnotesize The $\alpha$ (left) and $\beta$ (right) parameters 
for $\Delta y=1$ as a function of $N$.
The blue dots correspond to the data coming from the fit of 
our Ansatz
$P(v,\Delta y)$ to $\mathcal{P}$ for different values of $N$, 
while the green solid line is the constant fit in the large $N$ region.}
\label{fig.fitting_parameters}
\end{figure}

On the right panel of Fig.~\ref{fig.distribution}, a set of these PDFs
and their corresponding fits are shown for increasing values of $N=2^k$,
$k=3,6,\dots, 16$, and for finite $\Delta y=1$. We observe that the
peak position shifts linearly with $k$, i.e., increases logarithmically
with $N$. More importantly, the shape of the PDF approaches an
asymptotic form for increasing $N$. Indeed, this can be quantified by
studying the behavior of the fit parameters $\alpha$ and $\beta$ for
increasing $N$, which is shown in Fig.~\ref{fig.fitting_parameters} for
$\Delta y=1$.
After a linear increase with $\ln N$ for small $N$, they show a
convergent behavior for larger $N$, indicated by a clear flattening of
the curve for increasing $N$. We extract our estimates for the asymptotic 
values of $\alpha$
and $\beta$ in the large N limit from a fit to a constant in that flat
region\footnote{In a slight abuse of notation, we use 
$\alpha$, $\beta$ and $v_0$ for both the $N$ depending parameters 
as well as their asymptotic expressions. Whether we are refering to one or the 
other should be clear from the context.}.
Unfortunately, the onset of that flat region depends on
$\Delta y$ and occurs at larger $N$ for increasing $\Delta y$. This
fact ultimately puts a limit on the accessible range of propagation
distances $\Delta x$.

In order to identify the flat region where the parameters $\alpha$ and
$\beta$ have settled, we may inspect the $N$ dependence by hand as in
Fig.~\ref{fig.fitting_parameters}. Alternatively, we can use the
one-loop test described at the end of Subsect.~\ref{sec:one-loopcomp},
requiring $N$ to be sufficiently large to identify the logarithmic
one-loop divergence with a precision of, say, more than 90\%. In
practice, we find that both methods yield results for a minimum number
of $N$ which agree with one another.

\begin{figure}[h]
\begin{minipage}{.48\textwidth}
\includegraphics[width=1.05\textwidth]{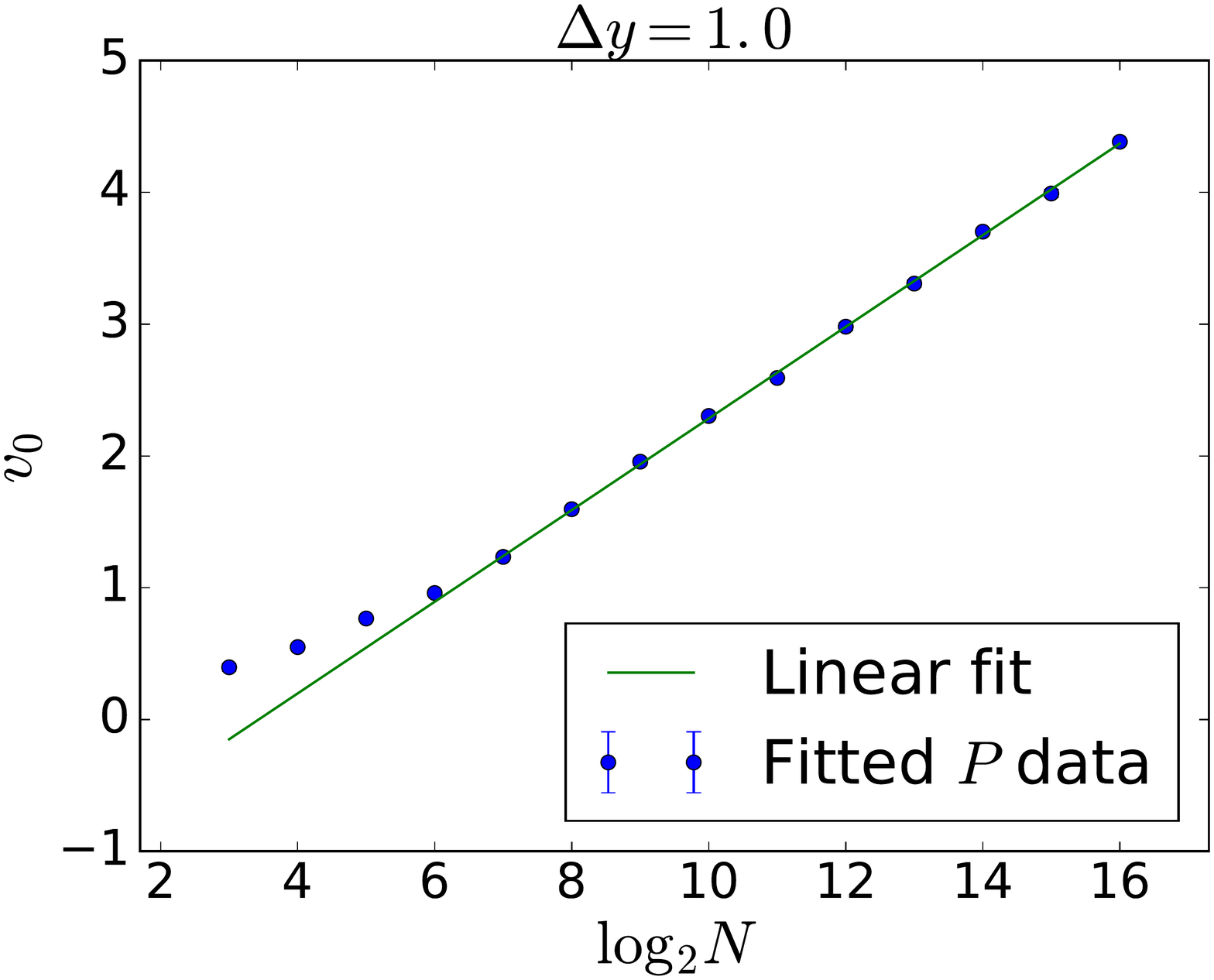}
\end{minipage}
\begin{minipage}{.48\textwidth}
\includegraphics[width=1.05\textwidth]{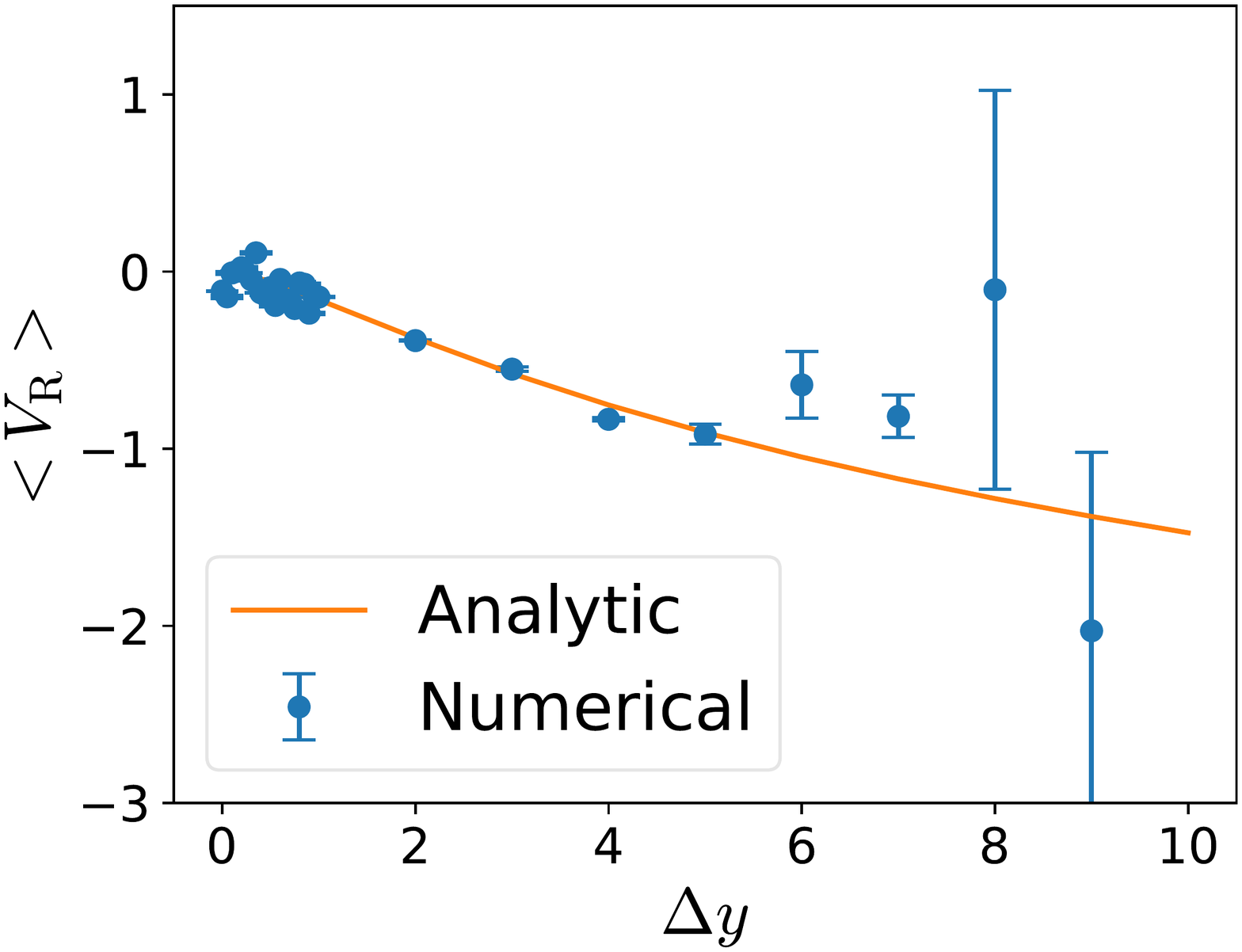}
\end{minipage}

\caption{\footnotesize Left: $N$ dependence of the fit parameter $v_0$ for $\Delta y=1$;
the blue dots show the results from fitting $P(v,\Delta y)$ to $\mathcal{P}$
for different values of $N$, 
and the green solid line represents the linear $\log_2 N$ fit in the large $N$ region. 
Right: self-consistency check using the regularized analytical 
result (eq. \eqref{eq.potential_1_loop}, solid orange line) and the
regularized numerical one (eq. \eqref{eq.V_R} using the large-$N$ fit values of
$\alpha$, $\beta$ and $b_{v_0}$, blue dots) for $T=1$.
}
\label{fig.media_v_parameters}
\end{figure}

As mentioned above, the log-divergence in $N$ is carried by the
parameter $v_0$. This is also obvious from the parametrization of the
expectation value of the interaction potential upon using the PDF fit
\eqref{pdf},
\begin{align}\label{pdf.mean_value}
\frac{1}{T} \left\langle
   V[y]\right\rangle_{y_I}^{y_F} \simeq\langle v \rangle_P := \int_0^{\infty} dv\, P(v)\,v = v_0+\frac{1+\alpha}{\beta}.
\end{align}
We have already worked out the  $\ln N$ divergence of the left-hand side
explicitly, whereas we have shown numerically that $\alpha$ and
$\beta$ converge to finite values for large $N$. Hence, the parameter
$v_0$ must behave as $v_0\sim(\ln N)/2$. As mentioned previously,
this is in agreement with the
shift of the peak of the PDF as visible in Fig.~\ref{fig.distribution}
(right panel) and can quantitatively be confirmed by an analysis of
the fit parameter results for $v_0$ -- see 
Fig. \ref{fig.media_v_parameters}, left plot,  where this behavior is depicted for $\Delta y =1$ 
together with a  large $N$ fit of the form
\begin{equation}
f_{v_0-\text{fit}}= a_{v_0} \log_2 N + b_{v_0}.
\label{eq:v0_largeNfit}
\end{equation}
In the same way as for the parameters $\alpha$ and $\beta$, the choice of the
large-$N$ fit region
can be done by either visual inspection or using the one-loop test.
Both methods provide equivalent results. 

These numerical results for $v_0$
are, however, not required for the following analysis. In fact, having
computed $\alpha$ and $\beta$ numerically, and knowing $\left\langle
   V[y] \right\rangle_{y_I}^{y_F}$ analytically from
\Eqref{eq:VdiscD4}, $v_0$ can be determined from \Eqref{pdf.mean_value},
\begin{align}\label{eq:def:v_0}
v_0=\frac{1}{T} \left\langle
   V[y]\right\rangle_{y_I}^{y_F} -\frac{1+\alpha}{\beta}.
\end{align}
This is the estimator we will use for $v_0$ in the following sections.
The advantage of this way of extracting $v_0$ is that the fully
available analytical information for the expectation value of the
interaction potential can be used. This facilitates at the same time
an exact subtraction of the log-divergencies in the course of the
renormalization procedure, see below. 

Computing $v_0$ numerically from the PDF fit instead serves as a worthwhile 
self-consistency check: if the obtained fits are valid, the WMC estimation
in formula \eqref{pdf.mean_value}
should hold true upon replacing the parameters $\alpha$, $\beta$ and $v_0$ by
their large-$N$ asymptotic expressions, while using 
the analytical expression  \eqref{eq:VdiscD4}  for  $\langle V\rangle_{\yI}^{\yF}$. 

To this end, in App. \ref{sec:largeN_asymptotic}
we determine the large-$N$ asymptotic expansion of 
\eqref{eq:VdiscD4}  up to $o(N^0)$. With this information, we can cancel the leading $\log_2 N$ 
contributions on both sides of the estimation \eqref{pdf.mean_value}. The result
is a ``renormalized'' self-interaction potential $V_{\text{R}}$. Our analytic
large-$N$ result is given in \Eqref{eq.potential_1_loop}, so that the corresponding numerical
estimate is given by
\begin{align}\label{eq.V_R}
\frac{1}{T} \left\langle
   V_{\text{R}}[y]\right\rangle_{y_I}^{y_F} \simeq b_{v_0}+\frac{1+\alpha}{\beta}-\frac{\gamma}{2}.
\end{align}
It should be clear that 
the attribute ``renormalized'' is justified, as the subtraction corresponds
precisely to the mass renormalization, cf. \Eqref{eq:mWS}. Note that also the
finite parts are subtracted, such that $\left\langle
   V_{\text{R}}[y]\right\rangle_{y_I}^{y_F} \to 0$ for $\Delta y\to
 0$. Results for $\left\langle
   V_{\text{R}}[y]\right\rangle_{y_I}^{y_F}$ are shown in the right panel of
Fig.~\ref{fig.media_v_parameters}. This demonstrates that the fit results are
indeed self-consistent up to distances of order $\Delta y\sim 5$.  This serves
as an indication for the region of confidence of our numerical
computations. We observe that the uncertainties in the numerical data, arising
from a propagation of uncertainties in the fit parameters, reflects the same
limitation, inasmuch as they strongly increase for distances near
$\Delta y\sim 5$.

In summary, the essential ingredient for obtaining non-perturbative
information about the propagator is the determination of the PDF
parameters $\alpha$ and $\beta$ as a function of the (rescaled)
distance $\Delta y$. In addition to using simulational data for
$\alpha$ and $\beta$ directly, we find it useful to introduce simple
fit functions for their distance dependence.

\begin{figure}[h]
\begin{minipage}{.48\textwidth}
\includegraphics[width=1.05\textwidth]{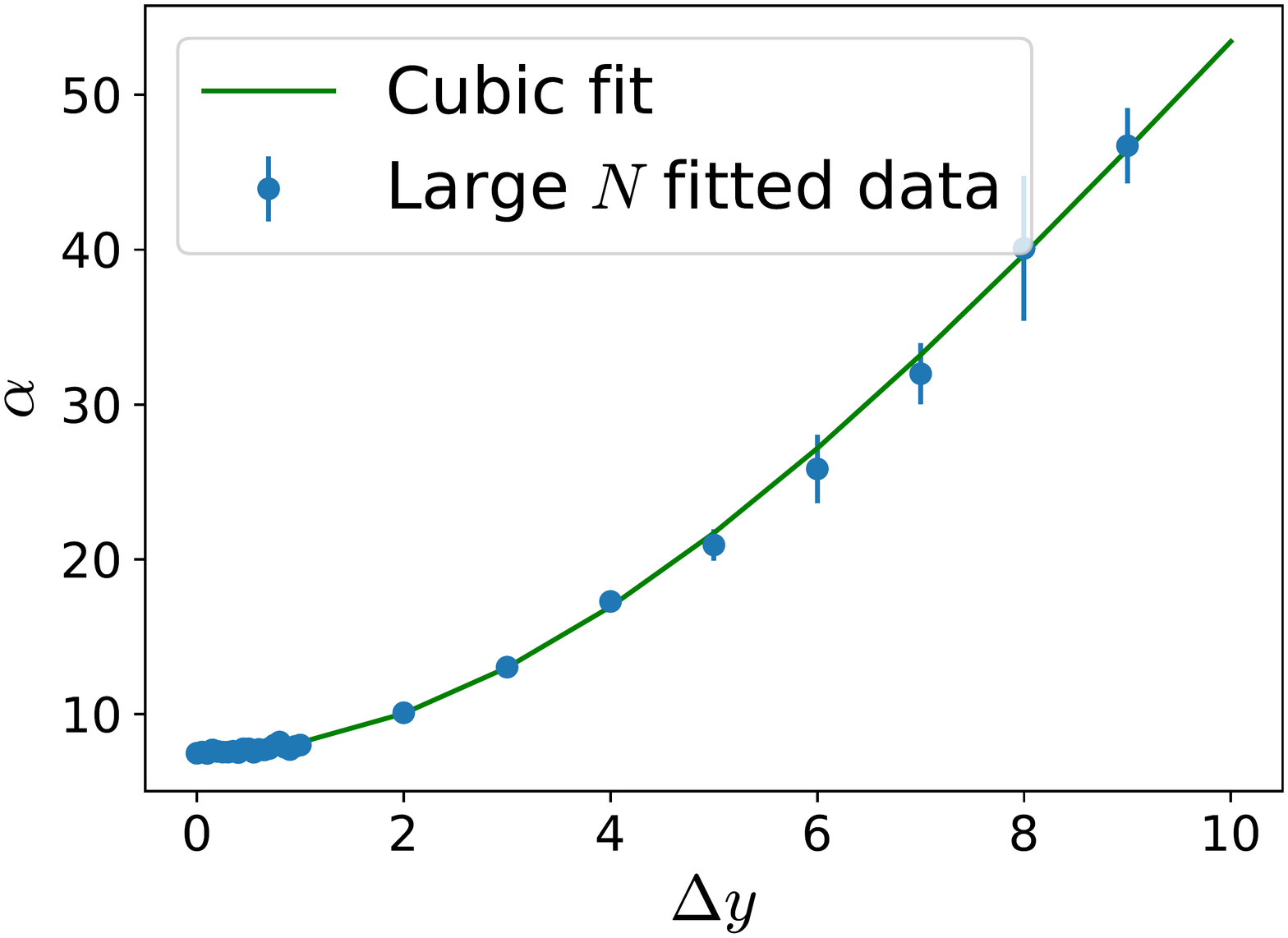}
\end{minipage}
\begin{minipage}{.48\textwidth}
\includegraphics[width=1.05\textwidth]{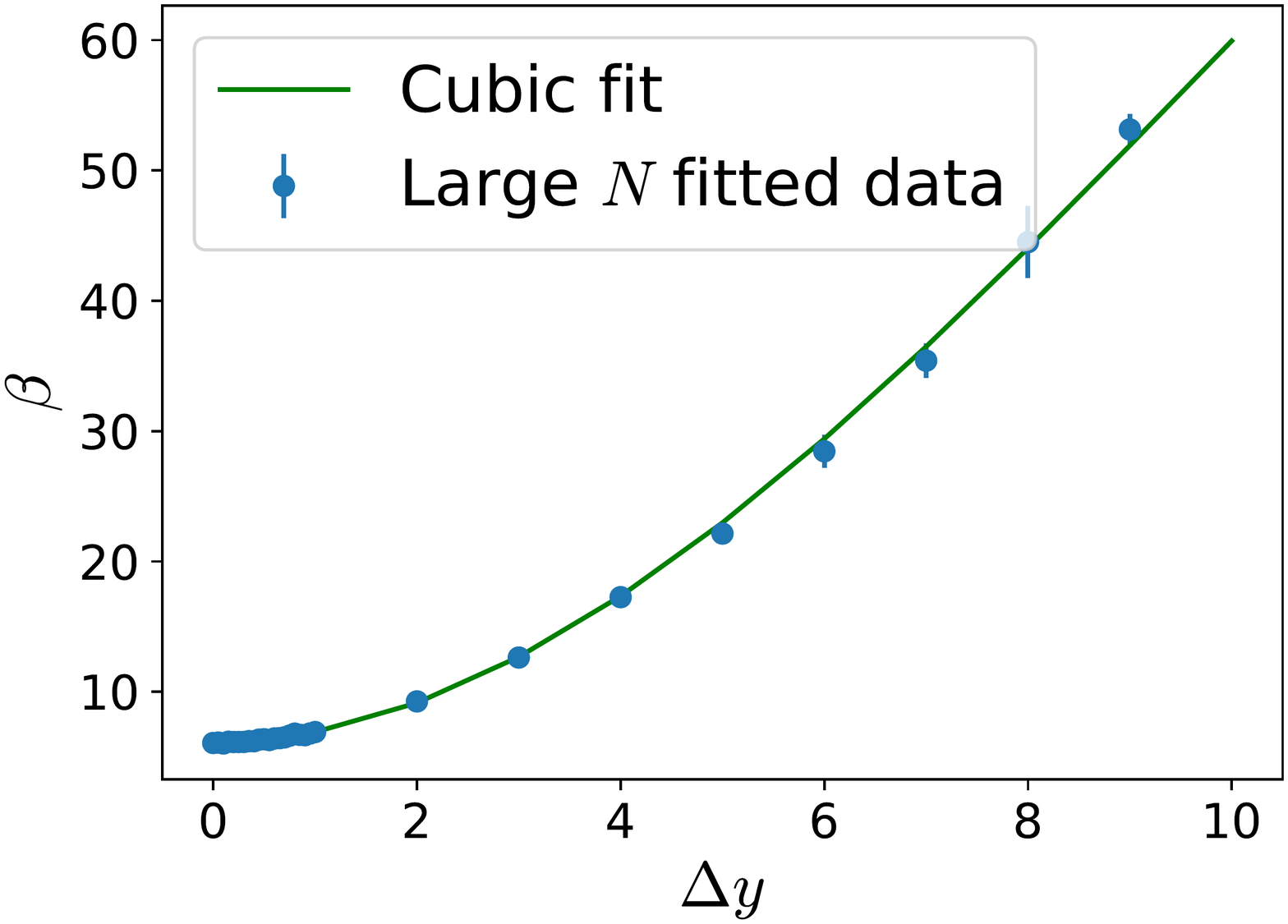}
\end{minipage}
\caption{\footnotesize PDF fit parameters $\alpha$ (left) and $\beta$ (right) as a function of the distance $\Delta y$.  The blue dots corresponds to the large $N$ fits of the $\alpha$ and $\beta$ parameters, while the green solid line is a third order polynomial fit.}
\label{fig.alpha_fitted}
\end{figure}

\begin{figure}[h]
\begin{minipage}{.48\textwidth}
\includegraphics[width=1.05\textwidth]{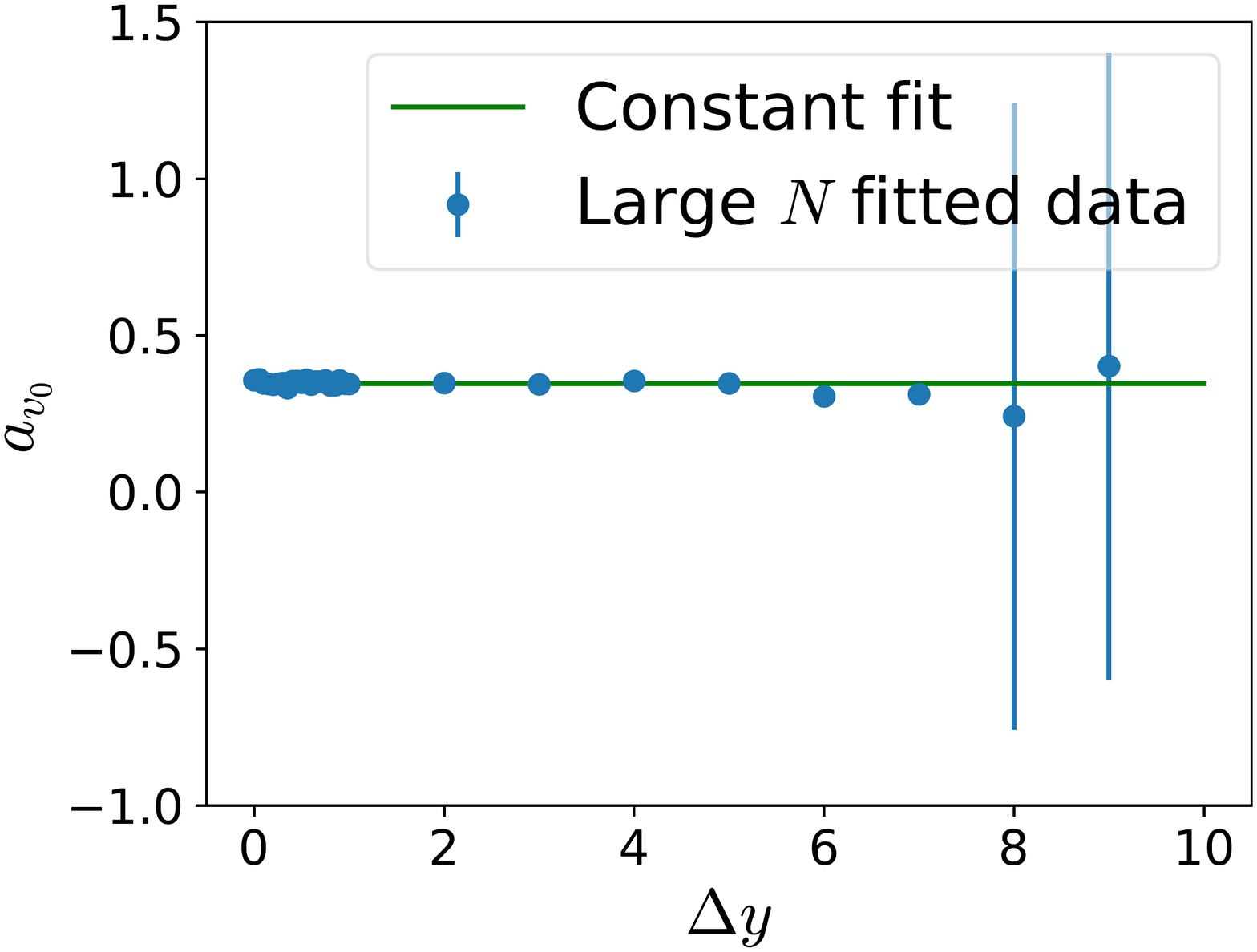}
\end{minipage}
\begin{minipage}{.48\textwidth}
\includegraphics[width=1.05\textwidth]{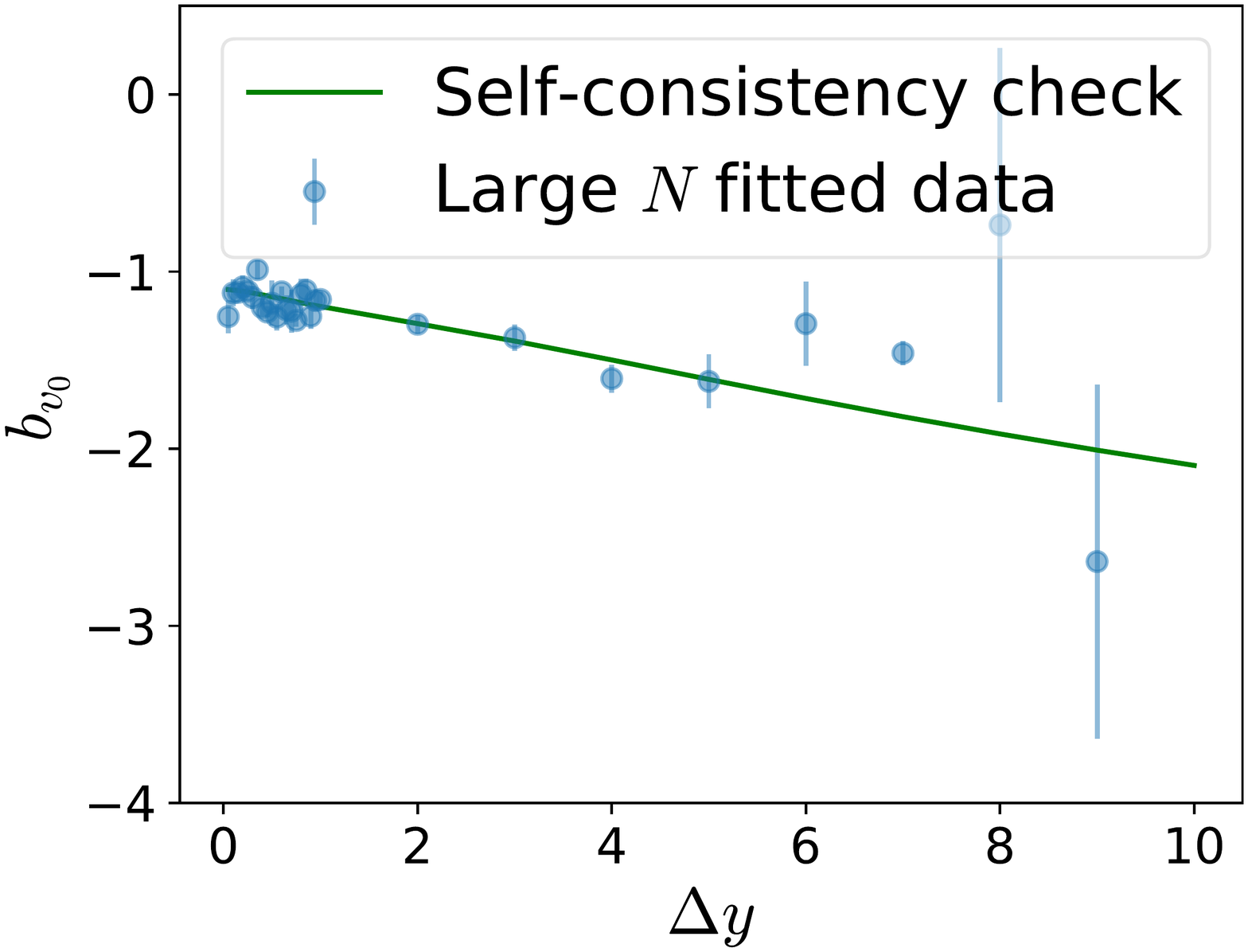}
\end{minipage}
\caption{\footnotesize PDF fit parameters $a_{v_0}$ (left) and and $b_{v_0}$
  (right) representing the fit parameter $v_0$ via eq. \eqref{eq:v0_largeNfit}
  as a function of the distance $\Delta y$. The blue dots corresponds to the
  large $N$ fits of the $v_0$ parameter. In the left panel, the green solid
  line is a constant fit,
  whereas in the rigth panel it shows the estimate for $b_{v_0}$ using
  \Eqref{eq:b_v_0}  as described in the text. }
\label{fig.v0_fitted}
\end{figure}

For completeness, we consider the distance dependence of all the parameters. 
On the one hand,  we find that both parameters $\alpha$ and $\beta$ are accurately
fitted by a polynomial of third degree in the a priori determined region of confidence $\Delta y \lesssim 5$,
while $a_{v_0}$ remains constant as predicted. The results of these fits are
\begin{align}\begin{split}\label{eq:fitted_parameters}
 \alpha &= 7.5116_8+ 0.6757_9 x^2-0.02164_3x^3,\\
  \beta &= 6.1276_2+ 0.8093_3 x^2-0.027124_8x^3,\\
  a_{v_0}&= 0.346749_2,
  \end{split}
\end{align}
as displayed in Fig. \ref{fig.alpha_fitted} 
for the $\alpha$ and $\beta$ parameters (left and right panel respectively), and in the left panel of 
Fig. \ref{fig.v0_fitted} for the slope $a_{v_0}$ of $v_0$.
The blue dots with error bars correspond to the large-$N$ asymptotic fits performed, whereas the 
green solid line depicts the fit functions of \eqref{eq:fitted_parameters}. We
observe that the coefficients of the cubic terms are significantly smaller
than the quadratic ones, suggesting a large radius of convergence of the
polynomial expansion. Also, the fits  \eqref{eq:fitted_parameters} describe
the data for $\alpha$ and $\beta$ well beyond the confidence region.

On the other hand, the $b_{v_0}$ parameter, characterizing the finite part of
$v_0$, exhibits a non-trivial behavior 
encoded in eq. \eqref{eq.potential_1_loop} and \eqref{eq.V_R}. 
Equation \eqref{eq:asymptotics_mean_V} in App.~\ref{sec:largeN_asymptotic}
suggests that a small distance fit of $b_{v_0}$ should require at least two parameters.
Instead, in order to avoid the proliferation of parameters, we test the quality of our
data by the following comparison: we start from eq. \eqref{eq.V_R},
solve it for $b_{v_0}$,
\begin{equation}
  b_{v_0}=\frac{\gamma}{2} - \frac{1+\alpha}{\beta} + \frac{1}{T} \left\langle
    V_{\text{R}}[y]\right\rangle_{y_I}^{y_F},
  \label{eq:b_v_0}
\end{equation}
and insert the fits \eqref{eq:fitted_parameters} of $\alpha$ and
$\beta$ on the right-hand side as well as the analytically determined form of
the renormalized self-interaction potential, cf. \Eqref{eq.potential_1_loop}. This
gives us a fully determined and analytically controlled estimator of $b_{v_0}$
without further parameters. We compare this result 
with the numerical data obtained by using the fit \eqref{eq:v0_largeNfit} 
in Fig.~\ref{fig.v0_fitted} (right)\footnote{The contribution from the
  uncertainties given in eq. \eqref{eq:fitted_parameters} for the $\alpha$ and $\beta$ parameters are smaller than the
  width of the plot line in Figure \ref{fig.v0_fitted} (right).}. We observe a
good agreement in the region of confidence $\Delta y\lesssim 5$. In fact,
$b_{v_0}$ is the only quantity parametrizing our data, where the noise beyond
the region of confidence shows up.

\subsection{Results for the propagator}\label{sec:fermion_propagator}

Let us now apply the PDF-based formalism to the worldline representation of
the propagator $G(\Delta x)$ for the charged scalar written in terms of the worldline expectation value 
of the exponential of the potential $V$, cf. eq. \eqref{eq:prop2} in Sect.
\ref{sec:worldline}:
\begin{align}\label{green.function}
 G(\Delta x)&= \frac{1}{(4\pi)^{d/2}} \int_{0}^{\infty} \frac{dT}{T^{d/2}} e^{-\mWS^2 T} e^{-\frac{\Delta x^2}{4T}} \left\langle e^{-\delta m^2T-g  V[y]}\right\rangle_{ \frac{\xI}{\sqrt{T}}}^{ \frac{\xF}{\sqrt{T}}}.
\end{align}
Using the ansatz for the PDF given by \eqref{pdf}, it is straightforward to
obtain an estimate of the expectation value in terms of the fit parameters,
\begin{align}\label{mean_value_exponential}
 \left\langle e^{-gV[y]}\right\rangle_{\Delta y}\simeq\left\langle e^{-g T v[y]}\right\rangle_{P,\Delta y}= F_{(\alpha(\Delta y), \beta(\Delta y))}(g T) e^{-g T v_0(\Delta y)},
\end{align}
where we have explicitly highlighted the dependence of the parameters on the
distance $\Delta y$ and defined the auxiliary function
\begin{align}
 F_{(\alpha, \beta)}(g T):= \left( \frac{\beta}{\beta+gT}\right)^{1+\alpha}.
\end{align}
%
%
Inserting eq.  \eqref{mean_value_exponential} into eq. \eqref{green.function}
leads us to
\begin{align}\label{green.function.final}
 G(\Delta x)&\simeq\frac{1}{(4\pi)^{d/2}} \int_{0}^{\infty} \frac{dT}{T^{d/2}} e^{-\mWS^2 T} e^{-\frac{\Delta x^2}{4T}} F_{(\alpha, \beta)}(g T) e^{-g T b_{v_0}+\frac{g}{2} T \gamma}\\
 &=:G_P(\Delta x).
\end{align}
As a quick check, consider the one-loop expansion of this formula:
performing a naive expansion of \eqref{green.function.final} in powers of $g$
leads to
\begin{align}
 G_P(\Delta x)&=\frac{1}{(4\pi)^{d/2}} \int_{0}^{\infty} \frac{dT}{T^{d/2}} e^{-\mWS T} e^{-\frac{\Delta x^2}{4T}} \left[1-\left(b_{v_0}+ \frac{1+\alpha}{\beta}-\frac{\gamma}{2}\right) g T+\cdots\right],
\end{align}
which together with eq. \eqref{eq.V_R} tells us that 
this is indeed the renormalized result corresponding to a linear expansion in the potential $V$. 

For a further analysis, it is useful to measure all dimensionful quantities in
units of the mass scale $\mWS$, and introduce
the dimensionless quantities
\begin{align}
 \bar{G}_P(\cdot)&=\frac{1}{\mWS^2}G_P(\cdot),\\
 \bar{g}&=\frac{g}{\mWS^2},\\
 \Delta \bar{x}&=\mWS \Delta x.
\end{align}
We also perform a corresponding rescaling of the propertime T with a
subsequent substitution by $T\to \Delta \bar{x} T$, yielding
\begin{align}\label{green.function.numerical}
\bar{G}_P(\Delta \bar{x})&=\frac{1}{(4\pi)^{2} \Delta \bar{x}^{2}} \int_0^{\infty} dT\, \mathcal{G}_{\Delta \bar{x},\bar{g}}(T ),
\end{align}
where the rescaled propertime integrand is
\begin{align}\label{eq:propagator.integrand}
\mathcal{G}_{\Delta \bar{x},\bar{g}}(T):&= \frac{1}{T^{2}} e^{-\left(1+\bar{g} b_{v_0}(\frac{1}{\sqrt{T}})-\frac{\bar{g}}{2}\gamma\right) \Delta \bar{x}^2 T} e^{-\frac{1}{4T}}  F_{\left(\alpha\left(\frac{1}{\sqrt{T}}\right), \beta\left(\frac{1}{\sqrt{T}}\right)\right)}(\Delta \bar{x}^2 \bar{g} T).
\end{align}

Formulas \eqref{green.function.numerical} and \eqref{eq:propagator.integrand}
assume that the values of the $\alpha$, $\beta$ and $b_{v_0}$ 
parameters are known for every positive argument.
However, we have already determined in Sect. \ref{sec:montecarlo} that our
region of confidence is limited to arguments $\Delta y$ satisfying 
$\Delta y\lesssim 5$. In the present rescaled form, this translates to
propertime values $T\gtrsim 0.04$. As the propertime is an integration
variable on the positive real domain, our result for the propagator can only be
considered a valid estimate if the integrand $\mathcal{G}_{\Delta
  \bar{x},\bar{g}}(\cdot)$ is localized in propertime regions satisfying this
constraint on $T>T_{\text{est}}=0.04$. 

Whether this constraint is satisfied depends on the 
coupling parameter $\bar{g}$ and the distance $\Delta\bar x$ under
consideration. The dominant features of the integrand arise from the interplay
between  two expontential terms: the first one with an exponent
proportional to $T$ stems from the mass term and controls the large-$T$
behavior, also relates to the long-range properties of the propagator. The
second exponential, with an exponent proportional to the inverse of T, controls
the small-$T$ (short-range) behavior. As a result, the propertime integrand
has a single peak, being exponentially damped to both sides of the peak. We
thus consider our estimate for the propagator reliable, as long
as the dominant part of the peak of the integrand is located at propertime
values satisfying $T>T_{\text{est}}$.

\begin{figure}[h]
\begin{minipage}{.48\textwidth}
\includegraphics[width=1.05\textwidth]{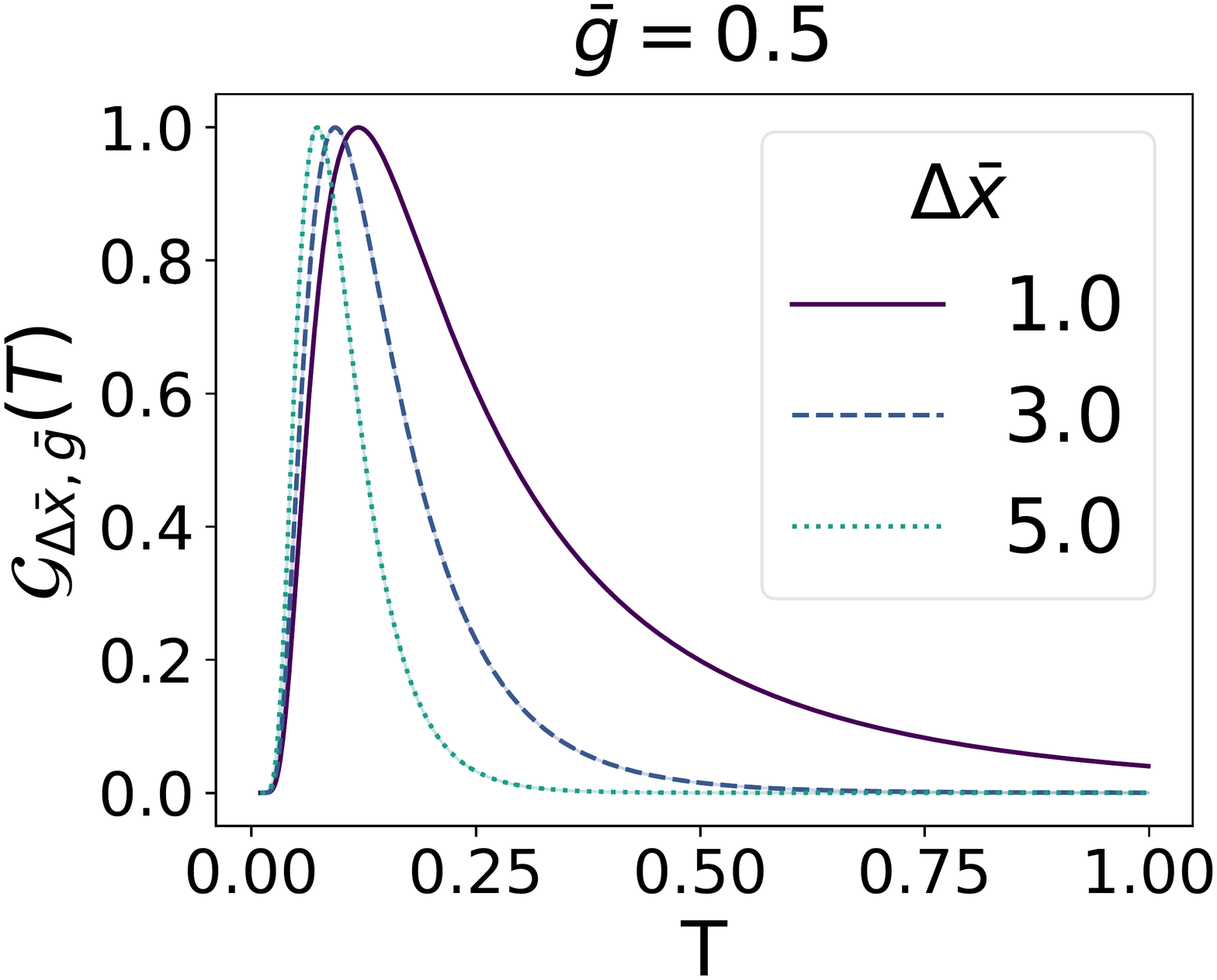}
\end{minipage}
\begin{minipage}{.48\textwidth}
\includegraphics[width=1.05\textwidth]{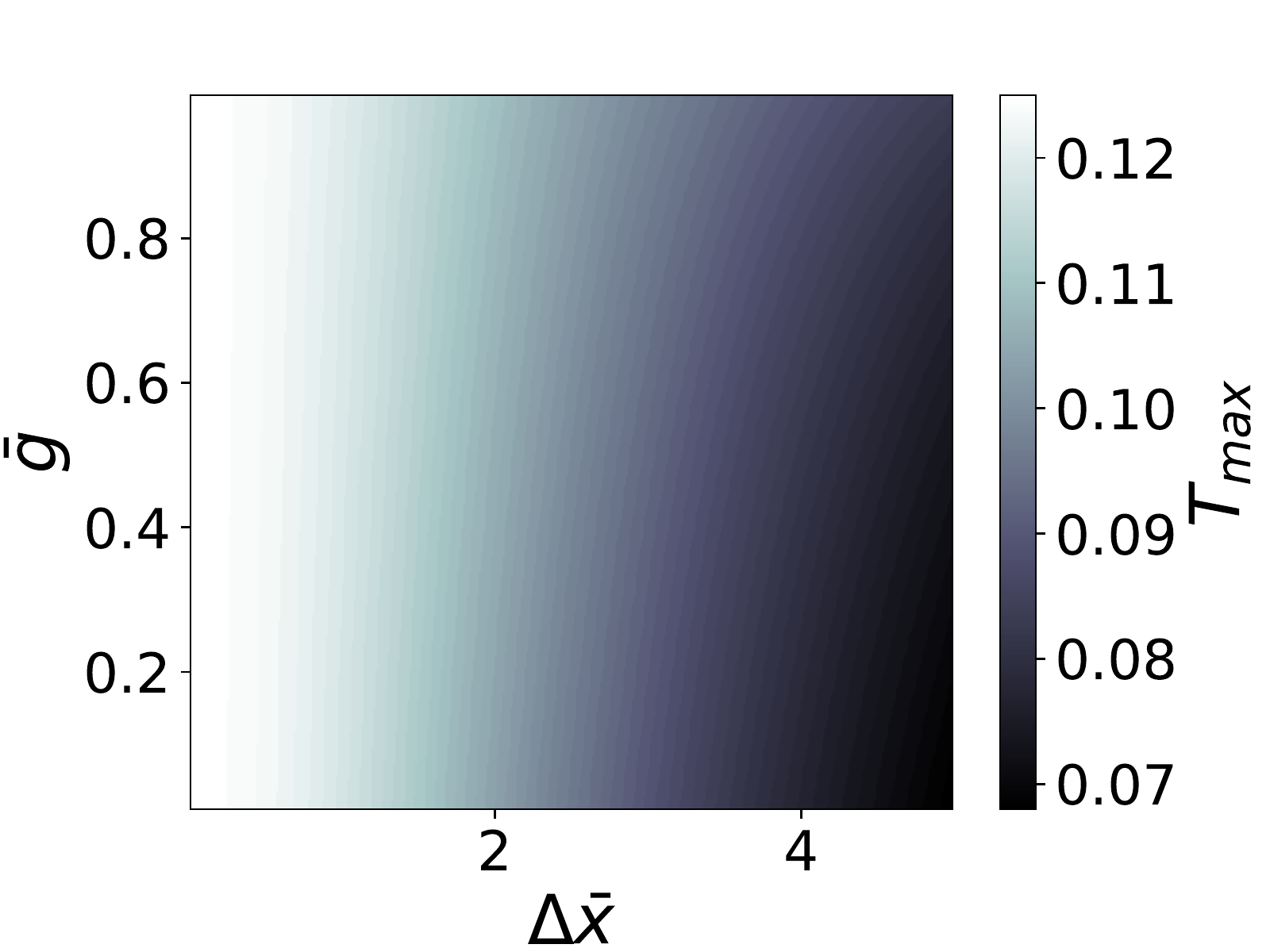}
\end{minipage}
\caption{\footnotesize Left: peak-normalized integrand $\mathcal{G}$ of
  \Eqref{eq:propagator.integrand}, for $\bar{g}=0.5$  and distances $\Delta \bar{x}=1,3,5$ (solid violet,
  dashed blue  and dotted cyan lines respectively).  Right:  density plot  of the propertime peak
  position $T_{\text{max}}$ of $\mathcal{G}$ as a function of $\bar{g}$ and $\Delta \bar{x}$.}
\label{fig.region_of_interest}
\end{figure}

As an example, consider the left panel 
of Fig.~\ref{fig.region_of_interest}. Here, 
we plot the propertime integrand \eqref{eq:propagator.integrand} as a function of the proper 
time $T$ for $\bar{g}=0.5$ and different values of the distance $\Delta
\bar{x}$, using the numerical data; for a better comparability, we have
normalized the peak of the integrand to one. Notice also that the numerical uncertainties
coming from a propagation of errors in the parameter uncertainties, cf. \eqref{eq:fitted_parameters}
are in these cases smaller than the width of the plotted lines. As expected,
the integrand is peaked aroung a value $T_{\text{max}}$ that tends to zero as the distance $\Delta \bar{x}$
increases.
In this particular case, the conclusion is 
that our formula \eqref{green.function.numerical} represents a satisfactory
estimate up to distances of order one. As a way to quantify the systematic error of our method when computing the propagator,
we assign an uncertainty given by the value of an integral analogous to \eqref{green.function.numerical}
with the upper boundary replaced by the value 0.04. We believe that this
procedure yields rather conservative error bars. 

Additionally, we show in Figure \ref{fig.region_of_interest} (right) 
a density plot of the peak position $T_{\text{max}}$ of $\mathcal{G}_{\Delta
  \bar{x},\bar{g}}$ as a function of the dimensionless distance $\Delta \bar{x}$
and coupling $\bar{g}$. Obviously, the region of self-consistency
$T_{\text{max}}>T_{\text{est}}=0.04$ corresponds to small values of
$\Delta\bar{x}$; we observe hardly any restriction on 
the coupling $\bar{g}$ in its allowed region.

\begin{figure}[h]
\begin{minipage}{.48\textwidth}
\includegraphics[width=1.05\textwidth]{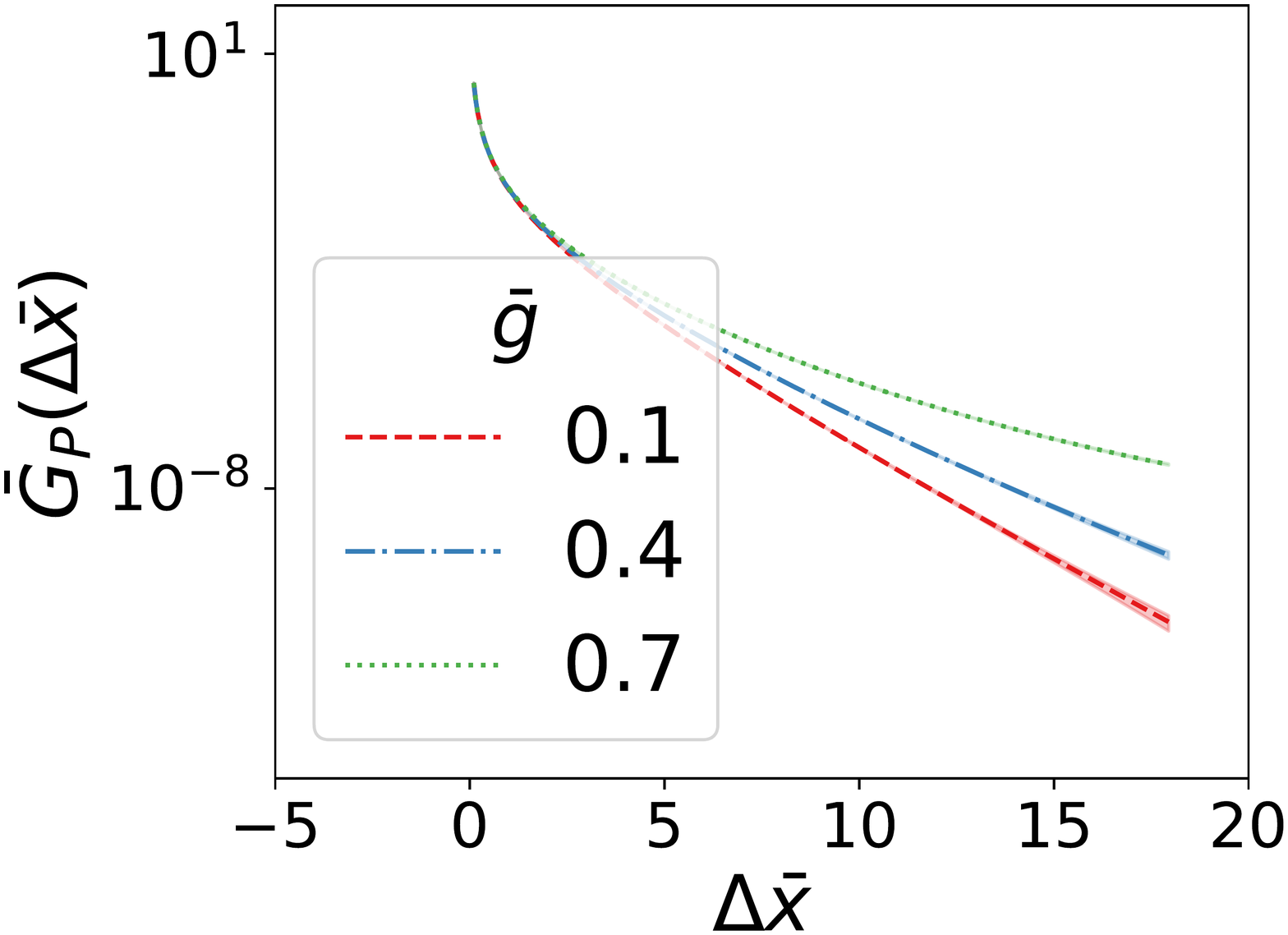}
\end{minipage}
\begin{minipage}{.48\textwidth}
\includegraphics[width=1.05\textwidth]{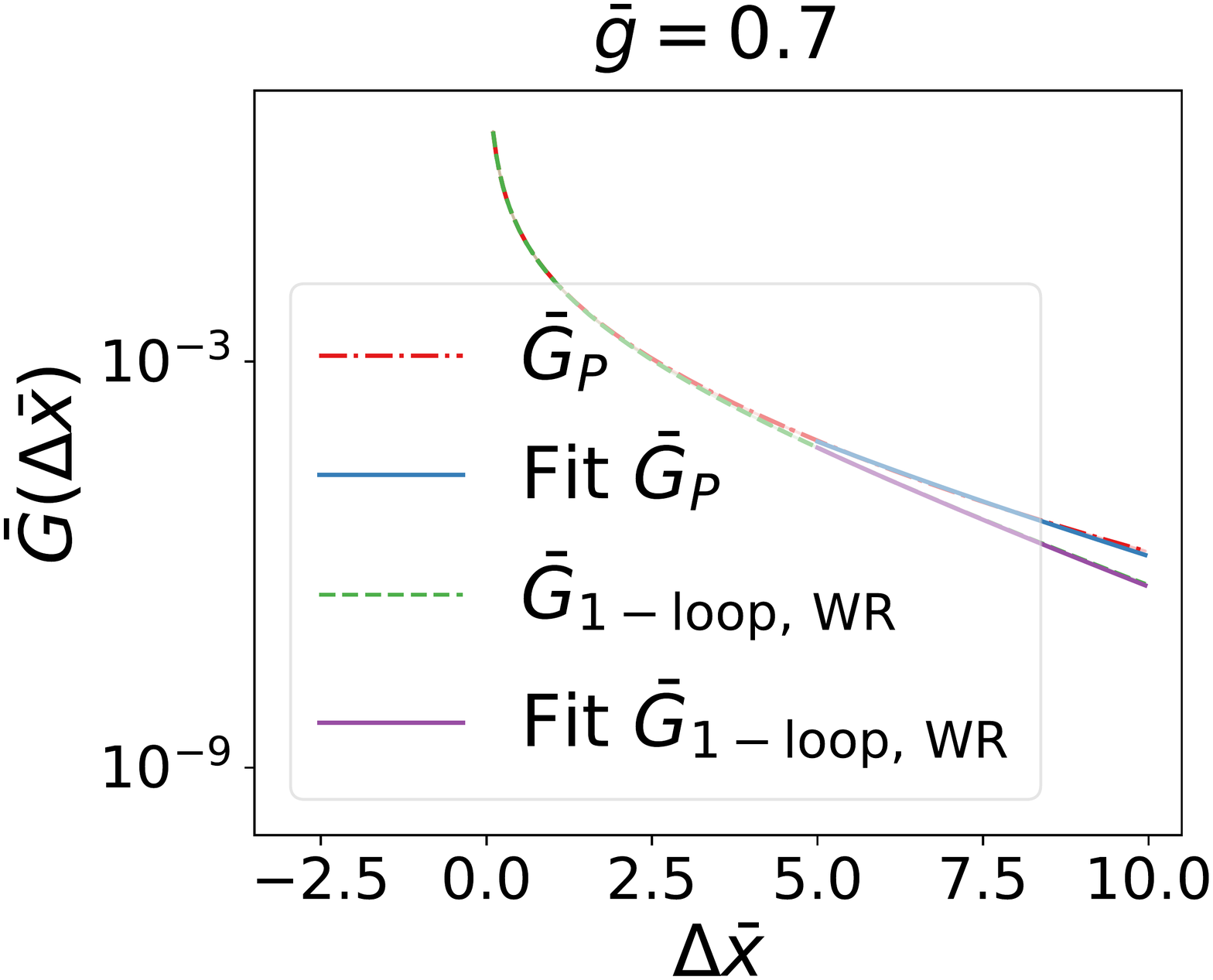}
\end{minipage}
\caption{\footnotesize Left panel: behaviour of the propagator
  $\bar{G}_{P}(\cdot)$  as a function of the distance for $\bar{g}=0.1$
  (dashed red line), $\bar{g}=0.4$ (dot-dahed blue line) and $\bar{g}=0.7$
  (dotted green line). Right panel: propagator (dot-dashed red line) and 
  the one-loop propagator $G_{1-\rm{loop},\,WR}$ in the worldline regularization (dashed green line)  
  together with their large distance fits (solid blue and violet line, for the 
  full and one-loop propagator, respectively).}
\label{fig.propagator}
\end{figure}

In fact, the region where we consider \eqref{eq:propagator.integrand} to
  be valid is limited by the constraint that the first exponential factor
  remains decaying for large $T$. This leads to a constraint on the coupling:
  \begin{equation}\label{eq.restriction_g}
    1+ \bar{g} b_{v_0}(0)-\bar{g}\frac{\gamma}{2} >0 \quad \Longrightarrow
    \quad
    \bar{g}<\bar{g}_{\text{c}}\simeq 0.72.
  \end{equation}
  For  couplings stronger than the critical one, the integrand \eqref{eq:propagator.integrand}
  becomes divergent for large propertimes. The maximum displayed on the
    right panel of Fig. \ref{fig.region_of_interest} therefore becomes only a
    local one beyond the critical coupling.
  
  Taken at face value, the limiting value $g=g_{\text{c}}$ corresponds to a
  coupling strength, where the propagator no longer decays exponentially for
  large distances, i.e. where the physical mass appears to tend to zero. At
  least in the present approximation, we are lead to conclude that the
  initially massive particle can become massless because of the dressing
  through the photon cloud if the coupling approaches the critical
  value. Whether this remains a feature of the model beyond our approximation
  is difficult to estimate, since it requires a careful study of the
  long-distance limit of the propagator.

This can be appreciated in the left panel of Fig. \ref{fig.propagator}, 
where the propagator is plotted as a function of the distance 
$\Delta \bar{x}$ for several values of the coupling. 
The exponential decay is indeed softened as the coupling increases. 
In the right panel of Fig. \ref{fig.propagator} we include, for $\bar{g}=0.7$, 
the comparison between the propagator and the  one-loop 
propagator $G_{\text{1-loop, WR}}$ in the worldline regularization 
(cf. eq. \eqref{eq:1_loop_propagator_WR} and in general App. \ref{sec:additional-1-loop} 
for its definition). Note that for large distances both of them 
decay exponentially as the free propagator given by expression \eqref{eq:propfree} does. 
For this reason we propose the following fit function 
for the propagators in the large distance regime,
\begin{align}
 f(x)=  \frac{A}{x^{3/2}} e^{-m^{\star}x},
\end{align}
where $A$ and $m^{\star}$, i.e. the physical mass corresponding to the pole
mass, are the fit parameters\footnote{This is indeed the first term in an
  asymptotic expansion of the one-loop propagator for large $x$, as proved in App.\ref{app:largedistance}.}. 

Remarkably, the fits\footnote{The fits are performed using the data for
  distances larger than $x=5$ which appears to be sufficiently deep in the
  asymptotic regime.} are in
excellent agreement with the propagators for $x$ larger than unity, as can be
seen in the right panel of Fig. \ref{fig.propagator}: the blue (violet)
solid line corresponds to the fit of the propagator $\bar{G}_{P}$
($G_{\text{1-\rm{loop}, WR}}$). Contrary to the free case, the physical mass in the
interacting case is not equal to the scale-setting mass $\mWS$, which is taken
as unity in these plots. Rephrasing this, in the light of the results
\eqref{eq:self_energy_app} and \eqref{eq:1_loop_propagator_WR}, it is
clear that the scale-setting mass $\mWS$ does in general not coincide with the
pole mass $m^{\star}$ of the propagator in complex momentum space.

\begin{figure}[h]
\begin{minipage}{.9\textwidth}
\begin{center}
\includegraphics[width=0.95\textwidth]{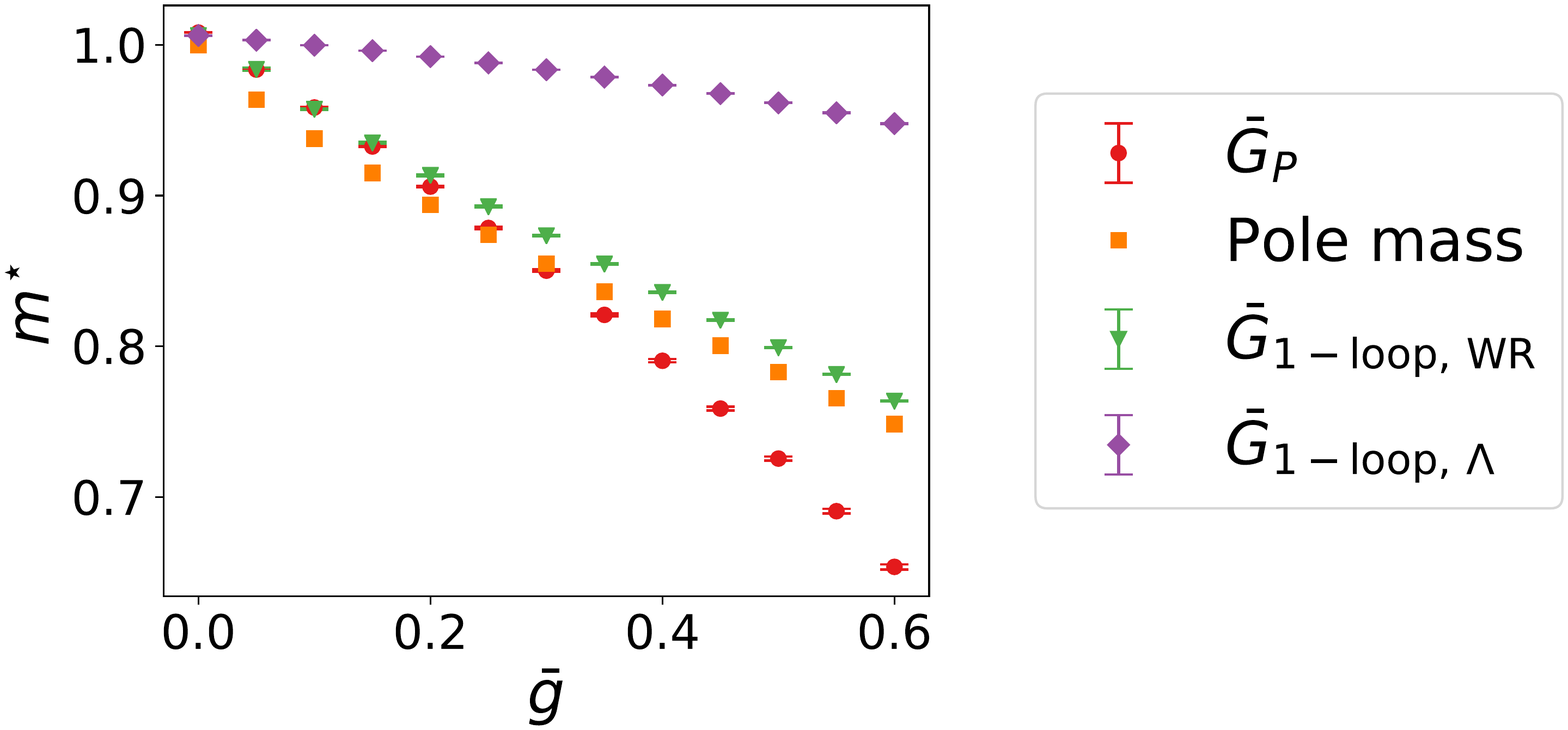} 
\end{center}

\end{minipage}
\caption{\footnotesize Pole mass $m^{\star}$ as a function of the coupling
  constant for various estimates: within worldline regularization, the
  analytical one-loop result is shown as orange squares, exhibiting
  satisfactory agreement with the same result extracted from a fit procedure
  (green triangles). The nonperturbative worldline result using the same fit
  procedure is shown as red  circles. For comparison, we show the one-loop
  pole mass extracted from the propagator using cutoff regularization (purple diamonds).}
\label{fig.propagator_smallx}
\end{figure}

Performing the analysis of physical masses for a wide range of coupling values, we summarize our
findings in Fig. \ref{fig.propagator_smallx}. As a first benchmark, we plot the
pole mass $m^{\star}$ as extracted  from the asymptotic expansion of the one-loop propagator
within the worldline regularization, cf. App. \ref{app:largedistance}, as orange
squares. Since our numerical data does not permit to go to asymptotically
large distances, we need to perform the numerical fits in a window of finite $x$ values
(we use $5\lesssim x \lesssim 10$). Applying this procedure to the one-loop
worldline result, we obtain the data points shown as green triangles. Over the
full range of couplings, this estimate is satisfactorily close to the
analytical result for the pole mass with deviations due to the fit procedure
on the few percent level.

We consider the smallness of these deviations as indicative for the reliability of the fit
procedure for the full propagator, whose results
are shown as red circles. A direct observation shows that the one-loop approximation is
quantitatively accurate up to $\bar{g}\lesssim 0.2$. For larger couplings, our
estimate for the pole mass of the  propagator decreases more rapidly than
the one-loop estimate. We interpret this as a consequence of the dressing of
the charged particle with the (scalar) photon cloud, which is described by the
infinite subclass of Feynman diagrams resummed by the nonperturbative
worldline formula \eqref{eq:prop2}. As discussed above, the charged particle even becomes
massless for $\bar{g}\to\bar{g}_{\text{c}}\simeq0.72$.

For further comparison, we also show the pole mass as derived from the
one-loop propagator using a momentum cutoff regularization
($\bar{G}_{1-\rm{loop},\,\Lambda}$, purple diamonds). The qualitative trend of
a decreasing pole mass for increasing couplings is also visible. However, the
dependence on the coupling appears much weaker. As emphasized above and
explained in more detail in App.~\ref{sec:additional-1-loop}, the
difference arises from the use of very different regularization
methods. Correspondingly, we expect the critical coupling $g_{\text{c}}$ to be
non-universal, i.e. to depend on the regularization scheme. Nevertheless, the
trend of the nonperturbative fluctuations to lower the pole mass for larger
couplings should persist in any regularization scheme.

Finally, we observe that the propagator in the deep UV region remains unaffected
by the fluctuations. For $\Delta \bar{x}\ll 1$, the behavior of both the
one-loop propagator $G_{1-\rm{loop},\, WR}$ and the propagator $\bar{G}_{P}$
coincide with that of the free propagator $\bar{G}_0$. This can be deduced
from the analytical expressions  \eqref{green.function.numerical} and
\eqref{eq:1_loop_propagator_WR} by expanding for small $\Delta \bar{x}$,
\begin{align}\label{eq:prop_small_x_expansion}
 \bar{G}_{1-\rm{loop},\,WR}(\Delta x)\sim \bar{G}_P (\Delta x) \sim \frac{1}{4\pi^2 \Delta x^2}, \qquad \Delta x\ll 1,
\end{align}
and is also confirmed by our numerical results. This observation is also in
line with the superrenormalizability of the model: the only possible
divergence is related to the mass operator -- UV-fluctuations are not strong
enough to also give rise to anomalous dimensions which could have modified the
short distance behavior given in formula \eqref{eq:prop_small_x_expansion}.

\section{Conclusions}\label{sec:conclusions}

We have extended nonperturbative worldline methods to a computation of a full
propagator in \sqed{}, a two-scalar model with cubic interaction.
For this, we have combined a compact worldline
representation for a large subclass of Feynman diagrams with information
carried by the probability distribution function of a relevant worldline
observable. This has enabled us to perform the renormalization of the model
nonperturbatively and to compute the propagator of the scalar electron for
large values of the coupling beyond the perturbative validity region.

The fully resummed subclass of diagrams is the dominant set in the formal
small flavor $\Nf\to 0$ limit. For the scalar electron propagator, it
diagrammatically corresponds to the electron line dressed by all possible
photon radiative corrections but without any additional electron loop. The
computation becomes accessible in the worldline formalism as it corresponds to
an expectation value of a worldline observable which we have been able to
compute using the method of probability distribution
functions. Algorithmically, we have followed corresponding earlier suggestions
for effective action computations \cite{Gies:2005sb} and generalizing these
methods for propagators on the basis of a newly developed $v$ lines algorithm,
cf. App.~\ref{sec:vlines}. These methods result in a semi-analytical
expression for the propagator, i.e. an analytical form that depends on a few
parameters to be determined from worldline simulations. Once the parameters
are computed numerically, the analysis of the propagator can be performed largely
analytically.

This gives rise to a number of concrete results for the model: as a general aspect, 
we observe that the propagator is always positive for the range of accessible
coupling values, so that we observe no violation of reflection
positivity. More precisely, 
we have analyzed the dependence of the propagator on the distance as a
function of the coupling. The large distance behavior is governed by the
physical (pole) mass corresponding to the inverse correlation length
characterizing the propagator.

For small couplings, our nonperturbative results coincide with that of the
one-loop approximation given by the resummation of the lowest-order
self-energy diagrams. Even though asymptotically large distances are
numerically difficult to deal with, the accessible distances already exhibit
the expected asymptotic behavior and allow for a determination of the pole
mass for comparatively large couplings $\bar{g}\sim0.5$. 

From about $\bar{g}\sim0.2$ on, we observe a clear deviation from the
perturbative estimate indicating the onset of a nonperturbative domain. The
inclusion of more diagrams corresponding to a full radiative dressing of the
electron with a photon cloud reduces the physical pole mass compared to the
leading-order perturbative estimate. Our results are compatible with the
existence of a critical coupling value for which the physical mass approaches
zero as a consequence of the radiative dressing.  This result agrees with the
fact that our semi-analytical expression \eqref{eq:propagator.integrand} shows
a constraint on the coupling parameter given by \eqref{eq.restriction_g},
which can be interpreted as an estimate of the critical coupling.

For these results, we have used a specific regularization prescription which
arises naturally in the worldline formalism in the form of a discretization of
dimensionless worldline trajectories in terms of polygons of $N$
segments. While this worldline regularization of keeping $N$ finite is simple and straightforward to
use in analytical as well as numerical worldline computations, the relation to
conventional regularizations of Feynman diagrams is more involved. This is
already obvious from the fact that the dimensionless parameter $N$ needs to be
related to a dimensionful parameter (such as a UV cutoff $\Lambda$ or a
renormalization scale $\mu$) which requires the dimensionality to be balanced
by the physical momentum or distance scale (at least in the deep Euclidean
region). As a consequence, our worldline result for the propagator determined with our
regularization differs from those of standard regularizations by computable
logarithmic terms. Such regularization dependences correspondingly occur for
all non-universal quantities such as the critical coupling value where the
mass vanishes. Still, the existence of a critical coupling is also suggested
by the behavior of the propagator in standard regularization schemes.

Finally, we observe that the small-distances behaviour of the propagator is not affected
by photonic corrections, neither perturbatively nor in the nonperturbatve
worldline computation. We consider this as evidence that the
superrenormalizable structure of the theory as suggested by power-counting is
preserved also nonperturbatively. As a result, the anomalous dimension of the
scalar electron field in \sqed{} remains zero both in perturbation theory
and beyond.

\section*{Acknowledgments}

The authors are grateful to Felix Karbstein, Horacio Falomir, Anton Ilderton and Luca Zambelli for helpful discussions. 
SAFV acknowledges support from the DAAD and the Ministerio de Educaci\'on de la Rep\'ublica Argentina 
under the ALE-ARG cooperation program.
This work has been funded by the Deutsche Forschungsgemeinschaft (DFG) under
Grant Nos. 416611371; 392856280 within the Research Unit FOR2783/1.

\appendix

\section{The $v$-lines algorithm}
\label{sec:vlines}

In the following, we construct an algorithm that generates open worldlines that 
obey a Gau\ss{}ian velocity distribution. This $v$ lines algorithm is a 
generalization of the efficient $v$ loops algorithm introduced in 
\cite{Gies:2003cv} which generates corresponding closed worldlines. 
Open worldlines can also efficiently be generated by a variant of the 
$d$ loop algorithm \cite{Gies:2005sb} that also works for open lines 
\cite{Schafer:2015wta,Schafer:2011as}, however, the following $v$ lines 
algorithm can be employed for an arbitrary number of points (for $d$ lines or 
loops they always come in powers of 2). 

First of all, we intend to create an ensemble of 
lines that run from $y_0$ to $y_{N}$ in $D$ dimensions
according to the discretized Gaussian velocity distribution
\begin{align}
\begin{split}
\bar{\mathcal{N}} \int_{y_0}^{y_N} \mathcal{D}y\, e^{-\frac{N}{4}\, S_W[y]} := \bar{\mathcal{N}}\int \prod_{j=1}^{N-1} d^Dy_j\, e^{-\frac{N}{4} \sum_{i=1}^{N} (y_i-y_{i-1})^2},
\end{split}\label{eq:gaussian_distribution}
\end{align}
where $\bar{\mathcal{N}}$ is the normalization needed to have a normalized-to-one distribution. 
The idea is to perform a set of linear variable transformations such
that the probability distribution becomes a Gau\ss{}ian one. As a first
step we complete the squares for $y_1$:
\begin{align}
 S_W=2\left(y_1-\frac{y_0+y_2}{2}\right)^2+\frac{1}{2}(y_2^2+y_0^2)-y_0\,y_2+\sum_{i=3}^{N} (y_i-y_{i-1})^2.
\end{align}
This naturally suggests to introduce a new variable $z_1$ defined by
\begin{align}
 z_1:=y_1-\frac{y_0+y_2}{2},
\end{align}
encoding all $y_1$ dependence of $Y$. The same procedure can be
applied to the dependence of the exponent on $y_2$:
\begin{align}
 S_W&=2\,z_1^2+\frac{3}{2}\left(y_2-\frac{y_0+2\,y_3}{3}\right)^2+\frac{1}{3}(y_3^2+y_0^2)-\frac{2}{3}y_0\,y_3+\sum_{i=4}^{N} (y_i-y_{i-1})^2.
 \end{align}
In this case, we obtain a purely quadratic dependence by introducing the new variable $z_2$,
\begin{align}
 z_2:=y_2-\frac{y_0+2\,y_3}{3}.
\end{align}

The general pattern for completing the squares for the $i$th variable $y_i$ is an expression of the form 
\begin{align}
 a_{i}y_i^2-2y_i(y_{i+1}+b_iy_0)=a_i\left(y_i-\frac{y_{i+1}+b_i y_0}{a_i}\right)^2-\frac{(y_{i+1}+b_i y_0)^2}{a_i},
\end{align}
with coefficients $a_i$ and $b_i$. After defining the variable $z_i$, we are left with the following $y_{i+1}$-dependent contributions:
\begin{align}
\left(2-\frac{1}{a_i}\right)y_{i+1}^2-2y_{i+1}\left(y_{i+2}+\frac{b_i}{a_i}y_0\right).
\end{align}
Consequently, the coefficients $a_i$ and $b_i$ are sequences that satisfy a system of recursion relations,
\begin{align}\label{v_lines_recursion}
 \left\lbrace\begin{array}{rll}
a_{i+1}&=2-\frac{1}{a_i}, &a_1=2,\\
 b_{i+1}&=\frac{b_i}{a_i}, &b_1=1.
 \end{array}\right.
\end{align}
The solution to these recursion relations can be straightforwardly obtained,
\begin{equation}
  \left\lbrace a_i=\frac{i+1}{i}, \quad b_i=\frac{1}{i} \right\rbrace,
\end{equation}
and hence, the general form of the variable $z_i$ reads
\begin{align}
  z_i=y_i-\frac{y_0}{i+1}-\frac{i}{i+1} y_{i+1}.
\end{align}
The quadratic form $Y$ rewritten in terms of these new $z_i$ variables
is finally diagonalized:
\begin{align}
 S_W=\sum_{i=1}^{N-1} \frac{i+1}{i}\,z_i^2+ c\,y_0^2+d\,y_N^2.
\end{align}
The values of the numbers $c$ and $d$, as well as the constant
Jacobian resulting from the change of variables $y_i\to z_i$ are not
relevant when computing expectation values, since they cancel with the
contributions coming from the corresponding normalization. Therefore we are left
with the task to generate a Gau\ss{}ian probability distribution for
the variables $z_i$, what is straightforward, e.g., with the
Box-M\"uller method.

In summary, the generation of $v$ lines with end points $y_0$ and $y_{N}$, and $N-1$ intermediate points 
obeying a Gau\ss{}ian velocity distribution can be performed as follows:
\begin{enumerate}
\item generate $N-1$ numbers $w_i, \, i=1\,\ldots,\,N-1$ via the Box-M\"uller method in such a way 
that they are distributed according to $e^{-w_i^2}$,
\item normalize the $w_i$, obtaining thus the auxiliary variables $z_i$:
  \begin{align}
    z_i=\sqrt{\frac{4}{N}}\sqrt{\frac{i}{i+1}}\,w_i;
  \end{align}
\item compute the points $y_{i}$ of the $v$ line for $i=N-1,\ldots,1$ by means of the recursive formula
 \begin{align}
  y_i=z_i+\frac{1}{i+1}y_0+\frac{i}{i+1}y_{i+1}.
 \end{align}
 
\end{enumerate}
For the special case of $y_0=y_N$, the algorithm generates closed $v$
loops attached at $y_0$ (so-called common point loops), which can be
transformed into common center-of-mass loops by a simple translation.



\subsection{Test of the $v$ lines algorithm}\label{sec:test-vlines}
As a way to test the code developed for the $v$ lines, we consider a simple 
model in which analytical expressions can be obtained. Consider then a probability 
distribution $P[y]$ for a discretized path $y$ in a $D$-dimensional space
given by  the ``action'' $S_W[y]$ in eq.\footnote{This would correspond to the probability that governs the 
behaviour of a quantum particle in a $D$-dimensional space, moving from $\yI$ to $\yF$ in a unit time $T=1$, 
and motivates the name action for $S_W[y]$. } 
\eqref{eq:gaussian_distribution}, i.e.
\begin{align}
\begin{split}
P[y]=\bar{\mathcal{N}} e^{-S_W[y]}.
\end{split}\label{eq:probability_particle}
\end{align}
Introducing an auxiliary variable $\kappa$ as a multiplicative factor in front
of the action,
we can straightforwardly compute expectation values of powers of the action, for example
\begin{align}\label{eq:mean_S}
 \begin{split}
 \left\langle S_W\right\rangle &=\bar{\mathcal{N}} \int dy_1 \cdots dy_{N-1} S_W[y] e^{-S_W[y]}\\
 &=\bar{\mathcal{N}} \int dy_1 \cdots dy_{N-1} \left. \frac{d e^{-\kappa S_W[y]}}{d\kappa}\right\vert_{\kappa=1}\\
 &=\frac{1}{2} (N-1) D + \frac{1}{4} \Delta y^2.
\end{split}\end{align}
Analogously, we can also compute the root mean square (RMS) of the action
\begin{align}
 \sqrt{\sigma^2(S_W)}= \sqrt{\left\langle S_W^2\right\rangle -\left\langle S_W\right\rangle^2} =\sqrt{\frac{(N-1)D}{2}}.
\end{align}
For the interested reader, the detailed computations can be followed in \cite{Gies:2005sb}.

Analytical results are compared to the corresponding numerical ones for the mean value of the action $S_W$ together with 
an error given by the RMS in Table\footnote{The values computed numerically with the $v$ lines algorithm 
are denoted with the subscript ``$\text{num}$''.} \ref{tab:average_Y} for several values of $N$ and distances $\Delta y=\yF-\yI$;
here, we have chosen an ensemble of $10^4$ lines and $D=4$. As can be seen,
the relative difference of these two values,
\begin{align}
\frac{\Delta S_W}{S_W}:= \frac{\left\langle S_W\right\rangle-\left\langle S_W\right\rangle_{\text{num}}}{\left\langle S_W\right\rangle},
\end{align}
remains smaller than the percent level even for a comparatively small number
of points per line as $N=32$.
Moreover, we see that also their RMSs are similar, and  overstimate the difference between 
the numerical and the analytical results in every case\footnote{We have intentionally kept
  non-relevant decimals for the uncertainties 
in Table \ref{tab:average_Y} in a non-standard fashion for reasons of illustration.}. 
\begin{center}
\begin{table}[h]
 \caption{Mean value of the action $S_W$, employing both the analytical result ($\left\langle S_W \right\rangle$)
 and the numerical computation involving the $v$ lines algorithm ($\left\langle S_W \right\rangle_{\text{num}}$),
 for different values of points per line $N$ and distances $\Delta y$, employing an ensemble of $10^4$ 
 lines and $D=4$. Their relative difference $\Delta S_W/S_W$ is also shown.}
 \begin{tabular}{| c c c c c|}
 \hline
 N  & $\Delta y$ & $\left\langle S_W \right\rangle_{\text{num}}$ & $\left\langle S_W \right\rangle$ & $\Delta S_W /S_W (\%)$\\
 \hline
 32  & 1 & $62.17_{7.96}$ & $62.25_{7.87}$ & 0.1 \\
 256  & 1 & $510.08_{23.5}$ & $510.25_{22.6}$ & 0.03\\
 2048 & 1 & $4094.66_{86}$ & $4094.25_{64}$ & -0.01\\
 256  & 10 & $535.14_{23.5}$ & $535.25_{22.6}$ & -0.03\\
 2048 & 10 &  $4117.55_{86}$ & $4119_{64}$ & 0.03\\
 \hline
 \end{tabular}
 \label{tab:average_Y}
 \end{table}
\end{center}

\begin{figure}[h]
\begin{minipage}{.48\textwidth}
\end{minipage}
\begin{minipage}{.48\textwidth}
\includegraphics[width=1.05\textwidth]{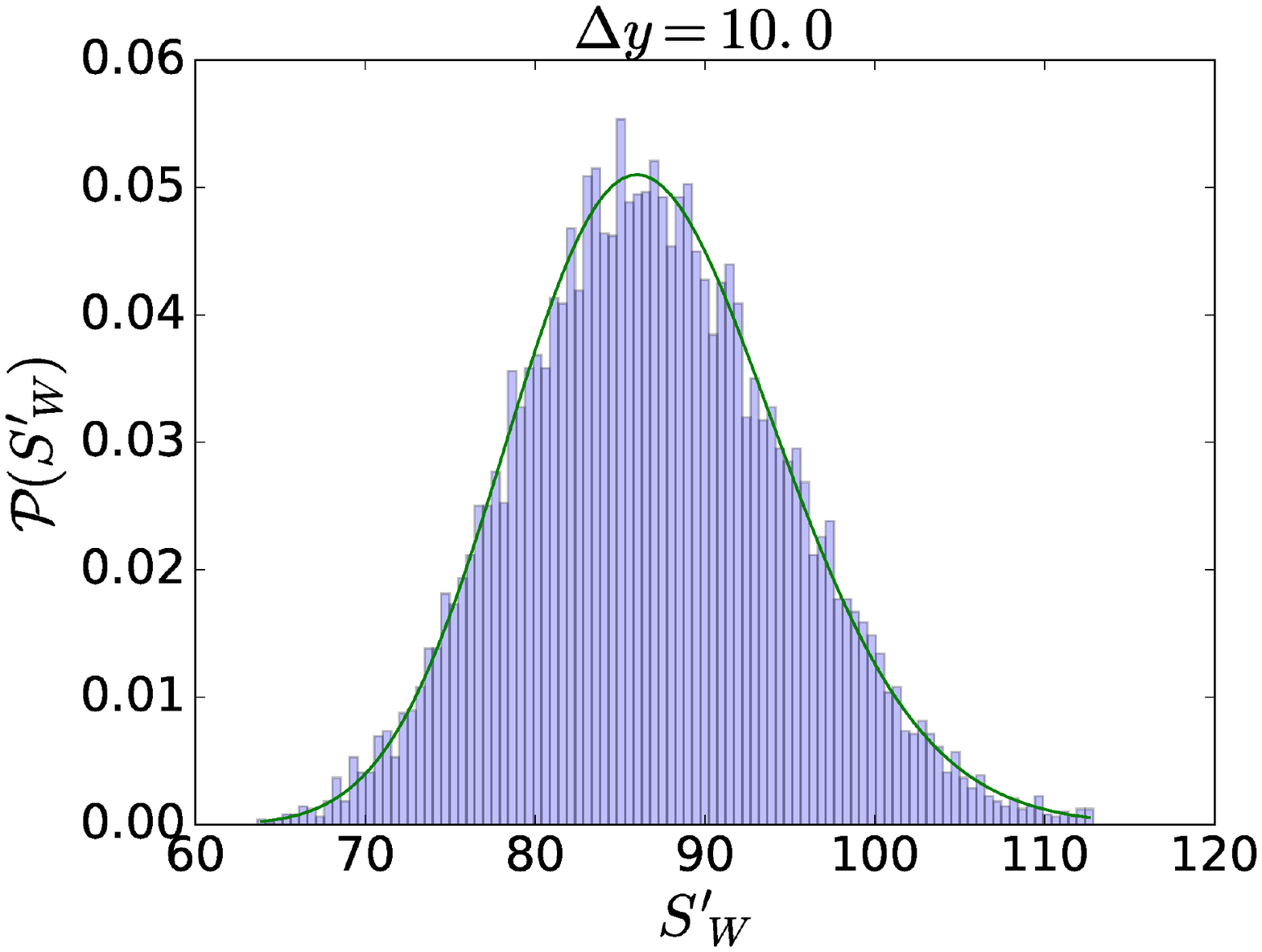}
\end{minipage}
\caption{\footnotesize Probability distribution function for the action $S_W$ considering 
$\Delta y=10$, $D=4$ and $N=2^5$. The histogram contains 100 bins and corresponds to an
ensemble of $10^4$ lines, whereas the solid green line is given by eq. \eqref{eq:distribution_action}.  }
\label{fig.histograms_action}
\end{figure}

 As a next step we consider the probability distribution 
 function $\mathcal{P}(S'_W)$ for the action $S_W$,
\begin{align}
\begin{split}\label{eq:distribution_action}
 \mathcal{P}(S'_W)&= \bar{\mathcal{N}} \int dy_1\cdots dy_{N-1} \delta(S_W-S_W') e^{-S_W}\\
 &= \mathcal{\bar{N}} (S'_W-S_{\text{class}})^{-1+\frac{(N+1)D}{2}} e^{-(S'_W-S_{\text{class}})} \theta(S'_w-S_{\text{class}}),
\end{split}
\end{align}
where the classical value for the action $S_{\text{class}}$ is given by
\begin{align}
 S_{\text{class}}= \Delta y^2.
\end{align}
The agreement between expression \eqref{eq:distribution_action} and the numerical data is remarkable
even for an ensemble of just $10^4$ lines. This can be seen from the histogram with 100 bins for $\Delta y=10$, 
$D=4$ and $N=2^5$, depicted in Fig. \ref{fig.histograms_action} as violet rectangles, and the corresponding analytic
expression (solid green line).

\section{One-loop contribution to the propagator in the discretized formulation}\label{sec:Vcomp}

As stated in Sect. \ref{sec:weak-coupling}, in order to obtain a closed expression 
for the mean value of the potential, we consider the Fourier transform of the interaction kernel
\begin{equation} 
\left\langle V[y] \right\rangle^{y_N}_{y_0}=\frac{8 T^{3-D/2}}{N^2 \Gamma \left( \frac{D-2}{2}\right)} \sum_{0=l<m}^{N-1} \mathcal{N}\int_{y_0}^{y_N}\mathcal{D}y\, e^{-\frac{N}{4} \sum_{j} (y_j-y_{j-1})^2} \int \frac{d^Dp}{(4\pi)^{D/2}} \frac{e^{ip(y_l-y_m)}}{p^2},
\end{equation}
which is evidently  of Gau\ss{}ian form and is valid for $D>2$. At this point, the method used in App. \ref{sec:vlines}
to diagonalize the quadratic form 
in the $v$ lines algorithm can be mutatis mutandis employed. Indeed, let us inspect the quadratic form
\begin{align}\label{Vcomp:quadratic}
Y=\sum_{j=1}^{N} \,(y_j-y_{j-1})^2-\frac{4i}{N}\,p\,(y_l-y_m).
\end{align}
Up to the ${(l-1)}$-th variable, the change of basis to the $z_j$ used before does the trick 
to diagonalize the quadratic form in the exponent, i.e.,
\begin{align}
z_j=y_j-\frac{y_0}{j+1}-\frac{j}{j+1} y_{j+1}, \quad 0<j<l.
\end{align}
The next term, however, receives an extra contribution coming from the interaction kernel
so that by completing the square for $y_l$ we obtain the contribution
\begin{align}
 a_l\left(y_l-\frac{b_l\,y_0+2ip/N+y_{l+1}}{a_l}\right)^2-\frac{(b_l\,y_0+2ip/N+y_{l+1})^2}{a_l}.
\end{align}

In turn, this implies that the structure of the remaining terms in the quadratic expression 
\eqref{Vcomp:quadratic} for $y_j$ is of the form 
\begin{align}
a_{j} y_{j}^2-2y_{j}\left(y_{j+1}+c_{j}\,\alpha_0\right),
\end{align}
where we have  restricted ourselves to $l<j<m$ and defined a shifted initial position $\alpha_0 :=b_ly_0+2ip/N$. 
Consequently  we are left with a pair of recursion relations analogous to \eqref{v_lines_recursion} 
with $c_j$ taking the r\^ole of the $b_j$, except for the fact that the initial condition is $c_l=1$. 
In other words, the initial condition corresponds in this case to a condition on the coefficients 
where the first potential insertion occurs. The diagonalizing variables are thus
\begin{align}
z_j=y_j-\frac{l}{j+1}\alpha_0-\frac{j}{j+1} y_{j+1} \text{, for } l\leq j < m.
\end{align}
The diagonalization proceeds analogously for the variables between the second insertion of the potential 
and the end of the line. The result is
\begin{align}
z_j=y_j-\frac{l}{j+1}\beta_0-\frac{j}{j+1} y_{j+1} \text{, for } m\leq j < N-1,
\end{align}
with the shifted initial position $\beta_0:=\frac{l}{m}\alpha_0-\frac{2ip}{N}$. 

Although the dependence of the quadratic form on the integrating variables $z_j$ 
is the same as in the free case, we are left with an extra factor coming from the 
completion of the squares which depends on the insertion points ($l$, $m$) of the potential 
and on the end points of the line. Notice also that the Jacobian for the change of variables 
$y_j\rightarrow z_j$ does not get modified, since 
the $z_j$ variables are only translated with respect to the ones defined in the free case. 
Ergo, the normalized expression is independent both of the Jacobian and of the 
determinant coming from the integration over the $z$ variables: 
\begin{align}\label{one_loop_potential}
\begin{split}
\left\langle V[y] \right\rangle^{\yF}_{\yI}&=\frac{8\, T^{3-D/2}}{N^2 \Gamma \left( \frac{D-2}{2}\right)} \sum_{m=1}^{N-1}\sum_{l=0}^{m-1}  \int \frac{d^Dp}{(4\pi)^{D/2}}  \frac{1}{p^2}e^{-p^2 \frac{(m-l)}{N}\left(1-\frac{(m-l)}{N}\right)-i\frac{(m-l)}{N}p (\yF-\yI)}.
\end{split}
\end{align} 
Notice that this expression depends only on the relative position $(m-l)$ of the insertions,
which could be understood as an inherited symmetry of paths ``translations'' in the continuum worldline expression. 

After performing the redefinition $(m-l) \rightarrow n$ of the summation index, we find the desired expression \eqref{eq:VdiscD}, exhibited in Sect. \ref{sec:weak-coupling}:
\begin{eqnarray} 
\left\langle V[y] \right\rangle^{\yF}_{\yI}&=&\frac{8\, T^{3-D/2}}{N^2 \Gamma \left( \frac{D-2}{2}\right)} \sum_{n=1}^{N-1} (N-n)  \int \frac{d^Dp}{(4\pi)^{D/2}}  \frac{1}{p^2}e^{-p^2 \frac{n}{N}\left(1-\frac{n}{N}\right)-i\frac{n}{N}p (\yF-\yI)}\nonumber\\
&=&2 \frac{T^{3-D/2}}{(\Delta y)^{D-2}} \sum_{n=1}^{N-1} \frac{N^{D-4} (N-n)}{n^{D-2}} 
\left[ 1- \frac{\Gamma\left( \frac{D-2}{2}, \frac{n \Delta y^2}{4 (N-n)} \right)}
{\Gamma\left( \frac{D-2}{2} \right)} \right].
\end{eqnarray}

\section{The large-$N$ asymptotics of the discretized $\langle V\rangle_{\yI}^{\yF} $  }\label{sec:largeN_asymptotic}
Recall that according to expression \eqref{eq:VdiscD4} the mean value  $\langle V\rangle_{\yI}^{\yF} $ 
of the potential in the discretized wordline regularization in $D=4$ is
given by
\begin{eqnarray}
\left\langle V[y] \right\rangle^{\yF}_{\yI}&=&2 \frac{T}{\Delta y^{2}} \sum_{n=1}^{N-1} \frac{ (N-n)}{n^{2}} 
\left[ 1- e^{- \frac{n \Delta y^2}{4 (N-n)}} \right].
\end{eqnarray}
In order to extract the large $N$ asymptotics out of this expression, we recast it in the following way:
\begin{align}
 \left\langle V[y] \right\rangle^{\yF}_{\yI}&=&2 \frac{T}{\Delta y^{2}} \sum_{n=1}^{N-1} \frac{ (N-n)}{n^{2}} 
\left[ 1- e^{- \frac{n \Delta y^2}{4 (N-n)}}- \frac{n \Delta y^2}{4 (N-n)} + \frac{n \Delta y^2}{4 (N-n)}\right].
\end{align}
The reason for this is that the sum of the last term can be explicitly computed as
\begin{align}
2 \frac{T}{\Delta y^{2}} \sum_{n=1}^{N-1} \frac{ (N-n)}{n^{2}}  \frac{n \Delta y^2}{4 (N-n)} = \frac{1}{2}T \left(\log N +\gamma\right).
\end{align}
On the other hand, it can be shown that the sum running over the first three terms can be replaced by an integral in the large-$N$ limit,
\begin{align}
\begin{split}
\left\langle V_{\text{R}}[y]\right\rangle_{\yI}^{\yF}:&=2 \frac{T}{\Delta y^{2}} \int_0^1 dn \frac{ (1-n)}{n^{2}} 
\left[ 1- e^{- \frac{n \Delta y^2}{4 (1-n)}}- \frac{n \Delta y^2}{4 (1-n)} \right]\\
&=2\frac{T}{\Delta y^2} \int^{\infty}_{0} \frac{dz }{z^2(1+z)} \left(1-\frac{z\Delta y^2}{4}-e^{-\frac{\Delta y^2z}{4}} \right) \label{eq.potential_1_loop}\\
&=\frac{T}{2 \Delta y^2} \left[4 e^{\frac{\Delta y^2}{4}} \text{Ei}\left(-\frac{\Delta y^2}{4}\right)+\Delta y^2-\left(\Delta y^2+4\right) \left(\gamma  + \log \left(\frac{\Delta y^2}{4}\right) \right) \right].
\end{split}\end{align}
This is what we call the renormalized mean value of the potential $\left\langle V_{\text{R}}[x] \right\rangle_{\yI}^{\yF}$.
From this expression, we can derive both a large and small $\Delta y$
asymptotics, reading
\begin{align}\label{eq:asymptotics_mean_V}
\left\langle V_{\text{R}}[y]\right\rangle_{\yI}^{\yF}&=\begin{cases}
 \frac{T\Delta y^2}{32}  \big(4 \log \Delta y+2 \gamma -3-4 \log 2\big) +\mathcal{O}(\Delta y^3), & \Delta y\ll 1\\[0.2cm]
 T\left(-\log \Delta y-\frac{\gamma}{2} +\frac{1}{2}+\log 2\right)+ \mathcal{O}(\Delta y^{-1}), & \Delta y\gg 1
 \end{cases}.
\end{align}

\section{Self-energy for the $\phi$ field}\label{sec:additional-1-loop}

In this appendix, we show how the worldline regularization and a cut-off 
regularization are related. First of all, recall that the one-loop propagator 
$G_{1-\rm{loop}} (x)$ may be expressed 
in terms of the self-energy $\Sigma(q)$ (the sunset Feynman diagram) as
\begin{align}\label{eq:1_loop_propagator}
  G_{1-\rm{loop}} (x)&= \int \frac{d^4q}{(2\pi)^{4}} \frac{e^{i x q}}{q^2+m^2-\Sigma(q)}.
\end{align}
Expanding to first order in the coupling $g$ and comparing with the expansion of the propagator \eqref{eq:prop2},
the self-energy contribution $\Sigma(q)$ in momentum space relates to a
worldline expectation value:
\begin{align}
\frac{\Sigma(q)}{(q^2+m^2)^2}&=\frac{1}{(4\pi)^{2}} \int d^4\Delta x\, e^{-iq\Delta x}  
\int_0^\infty \frac{dT}{T^{2}}\, e^{-m^2T} \, e^{-\frac{(\xF-\xI)^2}{4T}} 
                               \Big\langle -gV[x]\Big\rangle_{\xI}^{\xF}.
                               \label{eq:85}
\end{align}
Now, using the integral expression obtained in the second line of \eqref{eq.potential_1_loop} 
and ignoring momentarily the divergences that will be regularized later, we
get
\begin{align}\begin{split}
 \frac{\Sigma(q)}{(q^2+m^2)^2}
 &=-\frac{g}{8\pi^{2}}\int d^4\Delta x e^{-iq\Delta x}\int_{0}^{\infty} dTe^{-m^2 \Delta x^2 T-\frac{1}{4T}} \\
 &\hspace{4cm}\times\int^{\infty}_{0} \frac{dz }{z^2(1+z)} \left(1-e^{-\frac{z}{4T}} \right).
 \end{split}
 \end{align}
 The Fourier integral in this expression can be readily performed and, after
 rescaling the integration variables, the divergence for small $z$ becomes
 evident. As a regularization, we introduce a UV cutoff $\Lambda$ in units of
 mass,
 \begin{align}
 \begin{split}
\frac{\Sigma(q)}{(q^2+m^2)^2} 
 &=-\frac{g}{8} \frac{1}{(q^2+m^2)^2}\int_{0}^{\infty} \frac{dT}{T^2}   e^{-\frac{1}{4T}} \\
 &\hspace{3cm}\times \int^{\infty}_{\frac{m^2}{\Lambda^2}} \frac{dz }{z^2\left[ 1+z (q^2/m^2+1)\right]} \left(1-e^{-\frac{z}{4T}} \right).
 \end{split}
 \end{align}
The calculation of the remaining two integrals finally yields a closed
expression for the self-energy,
\begin{align}
 \begin{split}
\Sigma(q)
 &=\frac{g}{2} \left( 1+ \frac{m^2}{q^2}\right) \log (q^2/m^2+1) -\frac{g}{2} \log \frac{\Lambda^2}{m^2}\\
 &=:\Sigma_{\rm RS}(q) -\frac{g}{2} \log \frac{\Lambda^2}{m^2}.
 \label{eq:self_energy_app}
 \end{split}
\end{align}
Apart from giving a renormalized self-energy $\Sigma_{\rm RS}(q)$ equal to the one obtained in a 
standard computation in QFT (considering Feynman diagrams and using, e.g., dimensional regularization), 
eq. \eqref{eq:self_energy_app} also leads to a renormalization of the mass in
a way similar to the result of \eqref{eq:mWS}. This indicates that there should be a relation between 
the cut-off $\Lambda$ and the parameter $N$, which we will now heuristically explain. 

Since our worldline-regularized computation uses a discretization of the paths into a 
concatenation of  $N$ line segments, we are working with a resolution of
$\mathcal{O}(N^{-\frac{1}{2}})$ in configuration space or analogously, a 
resolution of order $\mathcal{O}(N^{\frac{1}{2}})$ in momentum space. 
It is then natural to assign to $N$ a relation of proportionality with the cutoff --
however, there should be a dimensionful energy scale characterizing the system and
linking them. In a loop expansion this r\^ole is played by the energy of the 
propagating particle, which in units of mass is $(1+q^2/m^2)$. 
Therefore, in order to do a meaningful comparison with a cutoff or dimensional regularization,
one needs to consider
\begin{align}
  \frac{\Lambda^2}{m^2}= \left(1+\frac{q^2}{m^2}\right)N\rightarrow \infty.
  \label{eq:N-Lambda}
\end{align}
Of course a thorough computation keeping track of $N$ from the beginning of the
computation gives the same result. Technically, the resolution in
  coordinate space contains  a propertime factor $\sqrt{T}$, as observed
  in Sect.~\ref{sec:weak-coupling}. This is linked to the energy scale of the
  worldline particle and hence causes the momentum dependent factor to appear
  on the RHS of \eqref{eq:N-Lambda}.

In summary, the renormalized expression $\Sigma_{\text{WR}}$ 
for the self-energy in the wordline regularization  is
\begin{align}\label{eq:sigma_WR}
 \Sigma_{\rm WR}(q)=&\frac{g}{2} \frac{\mWS^2}{q^2} \log (q^2/\mWS^2+1),
\end{align}
which is precisely the result for $\Sigma(p)$ when starting from
  \eqref{eq:85}, keeping track of the explicit dependence on a regularizing
  finite value of $N$ and renormalizing the mass a la \eqref{eq:mWS}.
Correspondingly, the one-loop propagator in the worldline regularization reads
\begin{align}\label{eq:1_loop_propagator_WR}
  G_{1-\rm{loop},\, WR} (x)&= \int \frac{d^4q}{(2\pi)^{4}} \frac{e^{i x q}}{q^2+m^2-\Sigma_{\text{WR}}(q)}.
\end{align}

\section{The large distance asymptotics of the one-loop propagator}\label{app:largedistance}

Let us analyze the large-distance asymptotics of the one-loop propagator. For
this, we start from expression \eqref{eq:1_loop_propagator_WR} for the
one-loop propagator, using the worldline regularized expression for the
self-energy given by \eqref{eq:sigma_WR}.  This can be rewritten as
\begin{align}
  G_{1-\rm{loop},\, WR} (x)
  &= \frac{1}{(2\pi)^{4}}\int dq_0 d^3q \, e^{i x q} \frac{q_0^2+q^2}{F(q^2+q_0^2)} ,
\end{align}
where we have chosen $x_0=0$ without loss of generality, and introduced the
function
\begin{align}
 F(q^2)=q^2(q^2+\mWS^2)-\frac{ g }{2} \mWS^2\log \left(\frac{q^2}{\mWS^2}+1\right).
\end{align}
The angular integration in the $q$ variable is straightforward, yielding
\begin{align}\label{eq:one_loop_integrand}
  G_{1-\rm{loop},\, WR} (x)
  &= \frac{1}{(2\pi)^{3}}\int_{-\infty}^{\infty}dq_0 \int_{-\infty}^{\infty}  dq 
  \frac{e^{i x q}}{ix}  \frac{q(q_0^2+q^2)}{F(q^2+q_0^2)} .
\end{align}
Now consider the integrand of eq. \eqref{eq:one_loop_integrand} in the complex $q$ plane. 
There are singularities for $q=\pm i (q_0^2+\mWS^2)$ (branch cuts arising from 
the logarithm) but there are also poles where $F(q^2+q_0^2)=0$. 
Let's call $q_{\star}^2=q_{\star}^2(g)$ the roots of the equation $F(-q^2_{\star})=0$. 
It is clear that whenever $g=0$ the roots of $F$ coincide with the branch points. 
However, once we turn on the interaction, two imaginary and complex conjugate roots appear. 
It can be proven that $q_{\star}^2<\mWS^2$, i.e., these poles are isolated and
not contained in the cut. 

At this point we are allowed to change the path over the real domain of $q$ into a path 
encircling the pole at $q=i \sqrt{q_0^2+q_{\star}^2}$ and a path going around the cut 
in the upper plane $\text{Im}(q)>0$. After this step, it is clear that the main 
contribution in the large $x$ limit is given by the integral around the pole. 
After a Laplace-type expansion of the remaining $q_0$ integral we obtain
our final expression
\begin{align}\label{eq:g_largex_expansion}
 G_{1-\rm{loop},\, WR} (x) \sim \frac{1}{(2\pi)^{3/2}}\frac{q_{\star}^{5/2} 
 \left(q_{\star}^2-\mWS^2\right)}{ \left(4 q_{\star}^4-6 \mWS^2 q_{\star}^2 +2 \mWS^4-g \mWS^2\right)} 
 \frac{ e^{-q_{\star} x}}{x^{3/2}}.
\end{align}
It is worth noticing the exponential decay of this expression and the power $x^{-3/2}$ 
of the accompanying prefactor. Also,
this expansion is only valid  for couplings $g< 2\, \mWS^2$, 
since at this point the roots of the function $F(q^2)$ become zero. 
This is also seen as a divergence of the prefactor in formula \eqref{eq:g_largex_expansion}.

\printbibliography

\end{document}